\PassOptionsToPackage{table,xcdraw}{xcolor}

\documentclass[sigconf]{acmart}

\AtBeginDocument{%
  \providecommand\BibTeX{{%
    \normalfont B\kern-0.5em{\scshape i\kern-0.25em b}\kern-0.8em\TeX}}}

\copyrightyear{2022}
\acmYear{2022}
\setcopyright{acmcopyright}
\acmConference[ICSE '22]{44th International Conference on Software Engineering}{May 21--29, 2022}{Pittsburgh, PA, USA}
\acmBooktitle{44th International Conference on Software Engineering (ICSE '22), May 21--29, 2022, Pittsburgh, PA, USA}
\acmPrice{15.00}
\acmDOI{10.1145/3510003.3510177}
\acmISBN{978-1-4503-9221-1/22/05}

\usepackage{booktabs}   
\usepackage{subcaption} 
\usepackage{array}
\usepackage{amsmath,amsfonts}
\usepackage{algorithm}
\usepackage[noend]{algpseudocode}
\usepackage{graphicx}
\usepackage{textcomp}
\usepackage{float} 
\usepackage{listings}
\usepackage{xspace}
\usepackage{multirow}
\usepackage{amsthm}

\usepackage{balance}

\usepackage[skins]{tcolorbox}

\usepackage{xcolor,pifont}
\newcommand*\colourcheck[1]{%
	\expandafter\newcommand\csname #1check\endcsname{\textcolor{#1}{\ding{52}}}%
}
\colourcheck{blue}
\colourcheck{green}
\colourcheck{red}

\newtcolorbox{myframe}[2][]{%
  enhanced,colback=white,colframe=black,coltitle=black,
  sharp corners,
  toprule=1.0pt,
  rightrule=0.3pt,
  leftrule=0pt,
  bottomrule=0pt,
  fonttitle=\itshape\scshape\large,
  left=0pt,right=5pt,top=5pt,bottom=3pt,
  attach boxed title to top right={yshift=-0.3\baselineskip-0.4pt,xshift=-5mm},
  boxed title style={tile,size=minimal,left=0.2mm,right=0.5mm,
    colback=white,before upper=\strut},
  title=#2,#1
}

\newcommand{\tool}{\textsc{DEAR}\xspace}

\newcolumntype{L}[1]{>{\raggedright\arraybackslash}p{#1}}

\newcommand{\code}[1]{{\footnotesize\texttt{#1}}}
\usepackage{amsthm}
 \definecolor{dkgreen}{rgb}{0,0.6,0}
\definecolor{gray}{rgb}{0.5,0.5,0.5}
\definecolor{mauve}{rgb}{0.58,0,0.82}
\begin{document}


\title[{\tool}: A Novel Deep Learning-based Approach for Automated Program Repair]{{\tool}: A Novel Deep Learning-based Approach\\ for Automated Program Repair}

\author{Yi Li}
\affiliation{
	\institution{New Jersey Inst. of Technology}
	\state{New Jersey}
	\country{USA}
}
\email{yl622@njit.edu}
\author{Shaohua Wang}
\authornote{Corresponding Author}
\affiliation{
	\institution{New Jersey Inst. of Technology}
	\state{New Jersey}
	\country{USA}
}
\email{davidsw@njit.edu}
\author{Tien N. Nguyen}
\affiliation{
	\institution{University of Texas at Dallas}
	\state{Texas}
	\country{USA}
}
\email{tien.n.nguyen@utdallas.edu}

\renewcommand{\shortauthors}{Yi Li, Shaohua Wang, and Tien N. Nguyen}












\begin{abstract}
The existing deep learning (DL)-based automated program repair (APR)
models are limited in fixing general software defects.
We present {\tool}, a DL-based approach that supports fixing for the
general bugs that require dependent changes at once to one or multiple
consecutive statements in one or multiple hunks of code.
We first design a novel fault localization (FL) technique for
multi-hunk, multi-statement fixes that combines traditional
spectrum-based (SB) FL with deep learning and data-flow analysis. It
takes the buggy statements returned by the SBFL model, detects the buggy
hunks to be fixed at once, and expands a buggy statement $s$ in a hunk
to include other suspicious statements around $s$.
%
%
%
We design a two-tier, tree-based LSTM model that incorporates cycle
training and uses a divide-and-conquer strategy to learn proper code
transformations for fixing multiple statements in the suitable fixing
context consisting of surrounding subtrees. We conducted several
experiments to evaluate {\tool} on three datasets: Defects4J (395
bugs), BigFix (+26k bugs), and CPatMiner (+44k bugs).  On Defects4J
dataset, {\tool} outperforms the baselines from 42\%--683\% in terms
of the number of auto-fixed bugs with~only the top-1 patches.  On
BigFix dataset, it fixes 31--145 more bugs than existing DL-based APR
models with the top-1 patches. On CPatMiner dataset, among 667 fixed
bugs, there are 169 (25.3\%) multi-hunk/multi-statement bugs. {\tool}
fixes 71 and 164 more bugs, including 52 and 61 more
multi-hunk/multi-statement bugs, than the state-of-the-art, DL-based
APR models.
%
%


\end{abstract}








\maketitle

\section{Introduction}


Researchers have proposed several approaches to help developers in
automatically identifying and fixing the defects in software. Such
approaches are referred to as {\em automated program~repair}
(APR). The APR approaches have been leveraging various techniques in
the areas of {\em search-based software engineering}, {\em software
  mining}, {\em machine learning (ML)}, and {\em deep learning (DL)}.

For {\em search-based
  approaches}~\cite{LeGoues-icse12,le2011genprog,martinez2016astor,qi2014strength},
a search strategy is performed in the space of potential solutions
produced by mutating the buggy code via operators. Other approaches
use software mining to {\em mine and learn fixing patterns} from prior
bug fixes
\cite{kim2013automatic,le2016history,liu2019avatar,tbar-issta19,nguyen2013semfix}
or similar code~\cite{icse10,ray-fse12}. Fixing patterns are at the
source code level \cite{liu2019avatar,tbar-issta19} or at the change
level~\cite{wen2018context,Simfix,koyuncu2018fixminer}. {\em Machine
  learning} has been used to mine fixing patterns and the candidate
fixes are ranked according to their
likelihoods~\cite{long2016automatic,long2017automatic,saha2017elixir}.
While some DL-based APR approaches learn similar
fixes~\cite{gupta2017deepfix,white2019sorting,white2016deep}, other
ones use machine translation or neural network models with
various~code~abstractions to generate
patches~\cite{chakrabortycodit,chen2018sequencer,hata2018learning,tufano2018empirical,see2017get,tufano2019learning,icse20}.

%

Despite their successes, the state-of-the-art DL-based APR
  approaches are still limited in fixing the {\em general defects, which
involve the fixing changes to multiple statements in the same or
different parts of a file or different files} (which are referred
to as {\em hunks}).
None of existing DL-based approaches can automatically fix the bug(s)
with dependent changes to multiple statements in multiple hunks
  at~once.
They supports fixing only individual statements. If we use
such a tool on the current statement, the tool treats that statement
as incorrect and treats the other statements as correct. This does not
hold since to fix the current statement, the remaining unfixed
statements must not be treated as correct code. Thus, it might
be~inaccurate when using existing DL-based APR tools to fix individual
statements for multi-hunk/multi-statement bugs.
While DL provides benefits for fix learning, this limitation
  makes the~DL-based APR approaches less capable than the other
  directions (search-based and pattern-based APR), which support
  multiple-statement fixes.

In this paper, we aim to advance deep learning-based APR
by~introducing {\tool}, a DL-based model that supports {\em fixing for the
general bugs with dependent changes at once to one or multiple
  buggy statements belonging to one or multiple buggy hunks of code}.
To do that, we make the following key technical contributions.

First, we develop a {\em fault localization (FL) technique for
  multi-hunk, multi-statement bugs that combines traditional
  spectrum-based FL (SBFL) with DL and data-flow analysis}. {\tool}
uses a SBFL method to identify the ranked list of suspicious buggy
statements. Then, it uses that list of buggy statements to {\em derive
  the buggy~hunks that need to be fixed together} by fine-tuning the
pre-trained BERT model~\cite{devlin2018bert}, to learn the
fixing-together relationships among statements.
We also design an expansion algorithm that takes a buggy statement $s$
in a hunk as a seed, and expands to include other suspicious
consecutive statements around $s$. To achieve that, we use an RNN
model to classify the statements as buggy or not, and use data-flow
analysis for adjustment and then form the buggy hunks.

Second, after the expansion step, we have identified all the buggy
hunk(s) with buggy statement(s). We develop a {\em compositional
  approach to learning and then generating multi-hunk, multi-statement
  fixes}. In our approach, from the buggy statements, we use a {\em
  divide-and-conquer strategy} to learn each subtree transformation in
Abstract Syntax Tree (AST). Specifically, we use an AST-based
differencing technique to derive the fine-grained, AST-based changes
and the mappings between buggy and fixed code in the training
data. Those fine-grained subtree mappings help our model avoid
incorrect alignments of buggy and fixed code, thus, is more accurate
in learning multiple AST subtree transformations of a fix.

Third, we have enhanced and orchestrated a tree-based, two-layer
Long Short-Term Memory (LSTM) model~\cite{icse20} with {\em
  an attention layer and a cycle training} to help {\tool} to learn
the proper code fixing changes in the suitable context of surrounding
code. For each buggy AST subtree identified by our fault localization,
we encode it as a vector representation and apply that LSTM model to
derive the fixed code. In the first layer, it learns the fixing
context, i.e., the code structures surrounding a buggy AST subtree. In
the second layer, it learns the code transformations to fix that buggy
subtree using the context as an additional weight.

Finally, there might be likely multiple buggy subtrees. To build the
surrounding context for each buggy subtree $B$, in training, we
include the AST subtrees {\em after} the fixes of the other buggy
subtrees (rather than those buggy subtrees themselves). The rationale
is that the subtrees after fixes actually represent the correct
surrounding code for $B$. (Note: in training, the fixed subtrees are
known).

We conducted experiments to evaluate {\tool} on three large datasets:
{\it Defects4J}~\cite{defects4j} (395 bugs), {\it
  BigFix}~\cite{icse20} (+26k bugs), and {\it CPatMiner
  dataset}~\cite{icse19-cpatminer} (+44k bugs). The baseline DL-based
approaches include DLFix~\cite{icse20},
CoCoNuT~\cite{lutellier2020coconut},
SequenceR~\cite{chen2018sequencer},
Tufano19~\cite{tufano2019learning}, CODIT~\cite{chakrabortycodit}, and
CURE~\cite{cure-icse21}.  {\tool} fixes 31\% (i.e., +11), 5.6\% (i.e.,
+41), and 9.3\% (i.e., +31) more bugs than the best-performing
baseline CURE on all three datasets, respectively, using only Top-1
patches and with seven times fewer training parameters on average. On
Defects4J, it outperforms those baselines from 42\%--683\% in terms of the
number of fixed bugs. On BigFix, it fixes 31–145 more bugs than those
baselines with the top-1 patches.
On CPatMiner, among 667 fixed bugs from {\tool}, there are 169
(25.3\%) multi-hunk/multi-statement ones. {\tool} fixes 71, 164, and
41 more bugs, including 52, 61, and 40 more multi-hunk/multi-statement
bugs, than existing DL-based APR tools CoCoNuT, DLFix, and CURE.
We also compared {\tool} against 8 state-of-the-art pattern-based APR
tools. Our results show that {\tool} generates comparable and
complementary results to the top pattern-based APR tools. On
Defects4J, {\tool} fixes 12 bugs (out of 47) including 7
multi-hunk/multi-statement bugs that the top pattern-based APR tool
could not fix.

In brief, the key contributions of this paper include




{\bf A. Advancing DL-based APR for general bugs with
  multi-hunk/multi-state\-ment fixes:} {\tool} advances DL-based APR
for general bugs. We show that DL-based APR can achieve the comparable
and complementary results as other APR directions.

  {\bf B. Advanced DL-based APR Techniques:}

  1) A {\em novel FL technique} for multi-hunk,
  multi-statement fixes that combines spectrum-based FL with DL and
  data-flow analysis;

  2) A {\em compositional approach} with a {\em divide-and-conquer strategy} to
  learn and generate multi-hunk, multi-statement fixes; and

  3) The design and orchestration of the two-layer LSTM model
  with the enhancements via the attention layer and cycle training.




{\bf C. Extensive Empirical Evaluation:} 1) {\tool} outperforms the existing
DL-based APR tools; 2) {\tool} is the first DL-based APR model
performing at the same level in terms of the number of fixed bugs as
the state-of-the-art, pattern-based tools and generate complementary
results; 3) Our data and tool are publicly available~\cite{AutoFix2021}.


\section{Motivation}
\label{sec:motiv}

\subsection{Motivating Example}
\label{motivexample}

\begin{figure}[t]
	\centering
	\renewcommand{\lstlistingname}{Method}
	\lstset{
		numbers=left,
		numberstyle= \tiny,
		keywordstyle= \color{blue!70},
		commentstyle= \color{red!50!green!50!blue!50},
		frame=shadowbox,
		rulesepcolor= \color{red!20!green!20!blue!20} ,
                xleftmargin=1.5em,xrightmargin=0em, aboveskip=1em,
		framexleftmargin=1.5em,
                numbersep= 5pt,
		language=Java,
                basicstyle=\scriptsize\ttfamily,
                numberstyle=\scriptsize\ttfamily,
                emphstyle=\bfseries,
                moredelim=**[is][\color{red}]{@}{@},
		escapeinside= {(*@}{@*)}
	}
	\begin{lstlisting}
public boolean verifyUserInfo(String UID, String password, String SSN) {
  String retrieved_password = "";
  String retrieved_SSN = "";
  if (UID != null) {
@-    retrieved_password = getPassword(UID);@
(*@{\color{cyan}{+\quad\quad retrieved\_password = getPassword(toUpperCase(UID));}@*)
(*@{\color{cyan}{+   \} else \{}@*)
(*@{\color{cyan}{+\quad \quad return false;}@*)
(*@{\color{cyan}{+ \}}@*)
@- boolean password_check= compare(password,retrieved_password);@
(*@{\color{cyan}{+  boolean password\_check= compare(passwordHash(password),retrieved\_password);}@*)
    if (password_check) {
      retrieved_SSN = getSSN(UID);
      boolean SSN_check = compare(SSN, retrieved_SSN);
      if (SSN_check) {
        return true;
      }
    }
    return false;
}
\end{lstlisting}
\vspace{-15pt}
\caption{A General Fix with Multiple Dependent Changes}
\label{fig:motiv}
\vspace{-15pt}
\end{figure}

Let us present a bug-fixing example and our observations for
motivation. Figure~\ref{fig:motiv} shows an example of a bug in
\code{verifyUserInfo}, which verifies the given user ID, password and
Social Security Number against users' records in the database. This
bug manifests in three folds. First, the developer forgot to handle
the case when \code{UID} is \code{null}. Thus, for fixing, (s)he added
an \code{else} branch at the lines 7--9.  Second, the developer forgot
to perform the uppercase conversion for the \code{UID}, causing an
error because the records for user IDs in the database all have
capital letters. The corresponding bug-fixing change is the addition
of the~call to \code{toUpperCase()} on \code{UID} at line 6. Third,
because the passwords stored in the database are encoded via hashing,
the input \code{password} from a user needs to be hashed~before it is
compared against the one in the database. Thus, the developer
added the call to \code{passwordHash()} on \code{password} before
calling the method \code{compare()} at line 11.  From this example, we
have the following observations:

\vspace{0.03in} \textbf{Observation 1 [A Fix with Dependent Changes to
    Multiple Statements]}: This bug requires the {\em dependent fixing
  changes to multiple statements at once in the same fix}: 1) adding
the \code{else} branch with the \code{return} statement (lines 7--9),
2) adding \code{toUpperCase} at line 6, and 3) adding
\code{passwordHash} at line 11. Making changes to the individual
statements one at a time would not fix the bug since both the given
arguments \code{UID} and \code{password} need to be properly
processed. \code{UID} needs to be null checked and capitalized, and
\code{password} needs to be hashed. Those dependent
changes~to~multiple statements must {\em occur at once in the same
  fix} for the program to pass the test cases.

The state-of-the-art DL-based APR
approaches~\cite{chen2018sequencer,icse20} {\em fix one individual
  statement at a time}. In Figure~\ref{fig:motiv}, the
fault localization tool returns two buggy lines: line 5 and line
10. Assume that such a DL-based APR tool is used to fix the statement
at line 5. It will make the fixing change to the statement at line 5
(e.g., modify line 5 and add lines 7--9), however, with the assumption
that the statement at line 10 and other lines are correct. With this
incorrect assumption, such a fix will not make the code pass the test
cases since both changes must be made.
Thus, the {\em individual-statement, DL-based APR tools cannot fix
  this bug by fixing one buggy statement at a time}. In general, {\em a
  bug might require dependent changes to multiple statements (in
  possibly multiple hunks) in the same fix}.

Moreover, the pattern-based APR tools might not be able to fix this
defect because the code in this example is project-specific and
might not match with any bug-fixing patterns.


%

\vspace{0.03in} \textbf{Observation 2 [Many-to-Many AST Subtree
    Transformations]:} A fix can involve the changes to {\em multiple}
subtrees. For~example, the \code{if} statement has a new \code{else}
branch. The argument of the call to \code{getPassword()} was modified
into the call to \code{toUpperCase()}. This fix also involves {\em
  many-to-many subtree transformations}. In this example, a fix
transforms the two buggy statements (line 5 and line 10), into four
statements (the \code{if} statement having new \code{else} branch, the
\code{return} statement at line 8, the modified statement with
\code{toUpperCase} at line 6, and the modified statement with
\code{passwordHash} at line 11). Thus, {\em a fix can be broken into
  multiple subtree transformations}, and if using a {\em
  composition approach with a divide-and-conquer strategy}, we can
learn the individual transformations.



\vspace{0.03in} \textbf{Observation 3 [Correct Fixing Context]:} {\em
  A bug fix often depends on the context of surrounding code}. For
example, to get the password from a given \code{UID}, one needs to
capitalize the ID, thus, in correct code, the method call to
\code{toUpperCase} is likely to appear when the method call to
\code{getPassword} is made. Therefore, {\em building correct fixing
  context is important}. In Figure~\ref{fig:motiv}, a model needs to
learn the fix (line 5 $\rightarrow$ line 6) w.r.t. surrounding code,
which needs to include the {\em fixed code} at line 11 (rather line 10
because line 10 is buggy). To fix line 5, the correct context must
include \code{passwordHash} at line~11. Thus, the correct context for
a fix to a buggy statement must include the fixed code of another
buggy statement $s$', rather than $s$' itself.

\subsection{Key Ideas}\label{keys}

From the observations, we draw the following key ideas:



\vspace{0.03in} \noindent {\bf \underline{Key Idea 1}. A Fault Localization Method
  for Multi-hunk, Multi-state\-ment Patches}: From Observation 1, we
design a novel FL method that {\em combines traditional spectrum-based
  FL (SBFL) with DL and data-flow analysis}. We use a SBFL to obtain a
ranked list of candidate statements to be fixed with their
suspiciousness scores. We extend the result from SBFL in two
tasks. First, we design a {\bf \em hunk-detection algorithm} to use DL
to detect the hunks that need {\em to be dependently changed
  together in the same patch}, because SBFL tool returns the
suspicious candidates for the fault, but not necessarily to be
fixed together. Second, we design {\bf \em an expansion algorithm}
that takes each of those detected fixing-together hunks and expands it
to include consecutive suspicious statements in the hunk.
%
In Figure~\ref{fig:motiv}, the SBFL tool returns line 5 as
suspicious.~After hunk detection, {\tool} uses data dependencies via 
variable \code{retrieve$\_$password} to include the statement
at line 10 as to be fixed as well.




\noindent {\bf \underline{Key Idea 2}. A Compositional Approach to
    Learning and Generating Multi-hunk, Multi-statement Fixes}:



\vspace{1pt}
{\em \underline{Divide-and-Conquer Strategy in Learning
    Multi-hunk/Multi-stmt} \underline{Fixes.}}  To auto-fix a bug with
multiple statements, a tool needs to make $m$-to-$n$ statement
changes, i.e., $m$ statements might generally become $n$ statements
after the fix. A naive approach would let a model learn the code
structure changes and make the alignment between the code before and
after the fix. Because a fix involves multiple subtree transformations
(Observation 2), during training, a model might incorrectly align the
code before and after the fix, thus, leading to incorrect learning of
the fix. For example, without this step, the model might map
\code{retrieved\_password} at line 10 to the same variable at line 6
(the correct map is line 11). Thus, to facilitate learning bug-fixing
code transformations, during training, we use {\em a
  divide-and-conquer strategy}. We integrate into {\tool} a
fine-grained AST-based change detection model to map the ASTs before and
after the fix. Such mappings enable {\tool} to learn the {\em more local fixing
changes} to subtrees.
For example, the fine-grained AST change detection can derive that the
statements at lines 4--5, and 7 become the statements at lines 4, and
6--9; and the statement at line 10 becomes the one at line 11. We can
break them into two groups and align the respective AST subtrees for
{\tool} to~learn.


\vspace{2pt}
{\em \underline{Compositional Approach in Fixing Multiple Subtrees}}.
We support the fixes having multiple statements in one or multiple
hunks by enhancing the design and orchestration of a tree-based LSTM
model~\cite{icse20} to {\em add an attention layer and cycle training}
(Section~\ref{model:sec}). While that model fixes one subtree at a
time, we need to enhance it to fix multiple AST subtrees at once.

Specifically, we modify its operations in the two-layers to consider
multiple buggy subtrees at once.
For example, during training, we mark each of the AST subtrees of the
statements at line 5 and line 10 before the fix as buggy. At the first
layer, for each subtree for a buggy statement, we {\em replace} it
with a pseudo-node, and consider the new AST with its (pseudo-)nodes
as the fixing context for the buggy statement. The pseudo-node is
computed via an embedding technique to capture the structure of the
buggy statement (Section~\ref{fixing-context}). At the second layer,
{\tool} learns the transformation from the subtree for the statement
at line 5 into the subtree for the fixed statements at the lines 6--9.
The vector for the fixing context learned from the first layer is used
as a weight in the code transformation learning in the second
layer. We repeat the same process for every buggy statement. For fixing, we
perform the composition of the fixing transformations for all buggy
statements at once.

\vspace{0.03in} \noindent {\bf \underline{Key Idea 3}. Transformation Learning with
    Correct Surrounding Fixing Context}: To learn the correct context
for a fix to a statement, we need to {\em train the
model} with {\em the fixed versions of the other buggy statements}
(Observation 3). For example, for training, to learn the fix to the
statement at line 5 with \code{toUpperCase}, a model needs to
integrate the fixed version of the other buggy line, i.e., the code at
line 11 with \code{passwordHash} as the fixing context (instead of the
buggy line 10). If the surrounding code before the fix is used
(i.e., line 10), the model will learn the incorrect context to fix the
line~5.

\subsection{Approach Overview}

\begin{figure}[t]
	\centering
        \includegraphics[width = 3.4in]{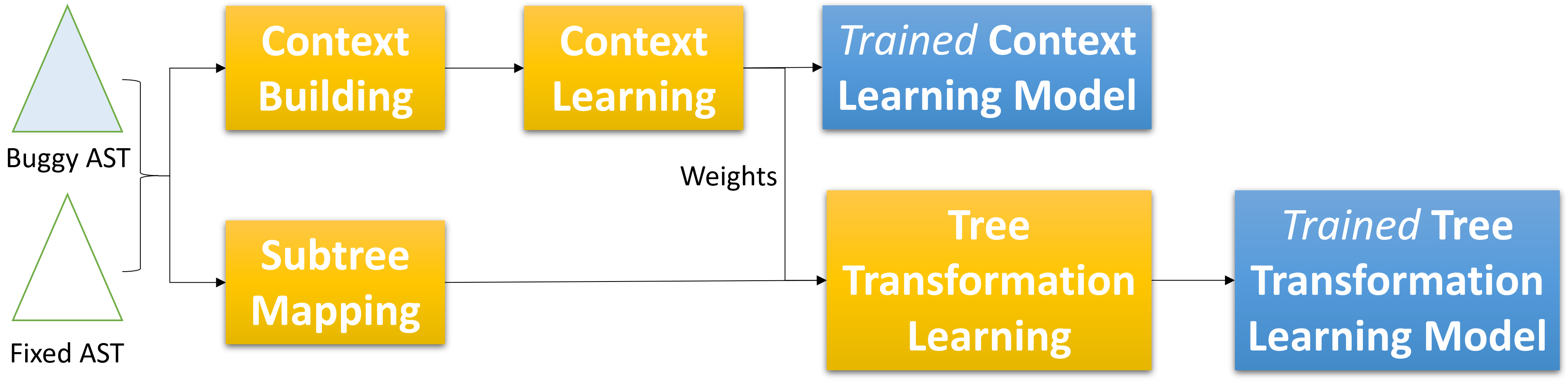}
        \vspace{-16pt}
        \caption{Training Process Overview}
	\label{Fig:train}
         \vspace{-6pt}
\end{figure}

\subsubsection{{\bf Training Process}}


The input for training includes the source code before and after a fix
(Figure~\ref{Fig:train}), which is parsed into ASTs. The output
includes the two {\em trained models for context learning} and {\em
for tree transformation learning} (fixing). The context learning model
(CTL) aims to learn the weights (representing the impact of the
context) to make an adjustment to the tree transformation learning
result. The tree transformation learning model (TTL) aims to learn
code transformation for the fix to a buggy AST subtree.

{\bf Context Learning}
(Sections~\ref{fixing-context}--\ref{model:sec}). The first step is to
build the before-/after-fixing contexts for training. With
divide-and-conquer strategy, we use CPatMiner~\cite{icse19-cpatminer}
to derive the {\em changed, inserted}, and {\em removed} subtrees (key
idea 2). As a result, the AST subtrees for the buggy statements are
mapped to the respective fixed subtrees. For each buggy subtree and
respective fixed subtree, we build two ASTs of the entire method as
contexts, one before and one after the fix, and use both of them for
training at the input layer and the output layer of the tree-based
LSTM context learning model (Section~\ref{model:sec}). To build the
correct context for each buggy subtree, we leverage key idea 3: we
train our model with the fixed versions of the other buggy
subtrees. Finally, the vectors computed from this learning are used as
the weights in tree transformation~learning.




{\bf Tree Transformation Learning}~(Section~\ref{codetrans:sec}).
We first use CPatMiner~\cite{icse19-cpatminer} to derive the subtree
mappings. To learn bug-fixing tree transformations, each buggy
subtree $T$ itself and its fixed subtree $T'$ after the fix are used
at the input layer and the output layer of the second tree-based LSTM for
training.
Moreover, the weight representing the context computed as the
vector in the context learning model is used an additional input in
this step. 



\subsubsection{{\bf Fixing Process}}

\begin{figure}[t]
	\centering
        \includegraphics[width =3.4in]{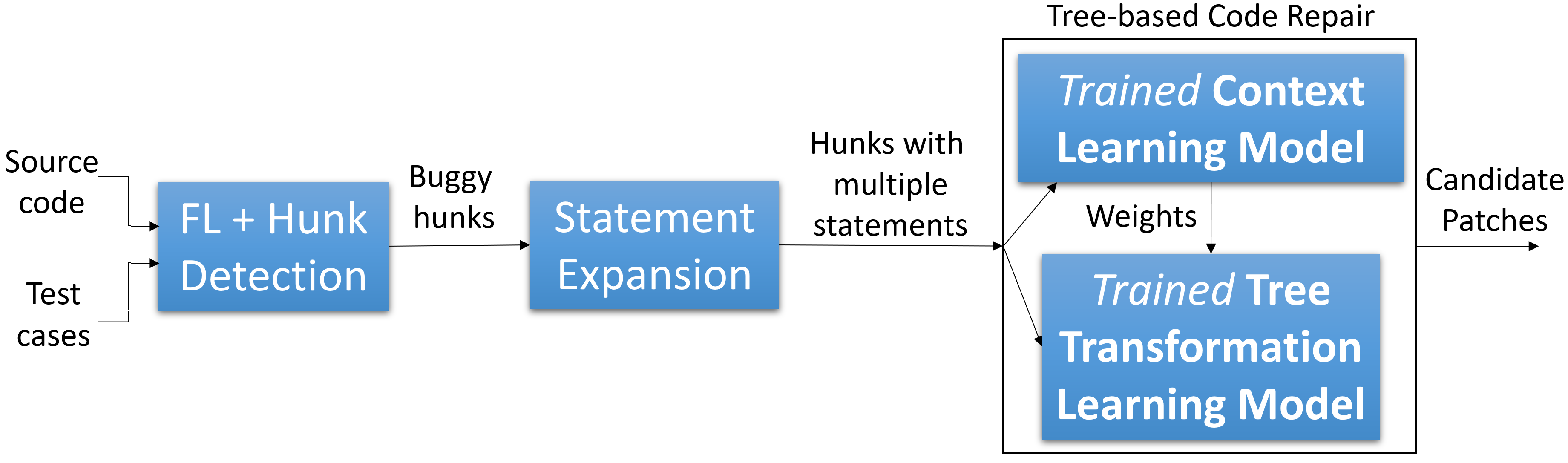}
       \vspace{-20pt}
	\caption{Fixing Process Overview}
	\label{Fig:test}
	\vspace{-10pt}
\end{figure}

Figure~\ref{Fig:test} illustrates the fixing process. The input includes
the buggy source code and the set of test cases.

{\bf Fault Localization and Buggy-Hunk Detection}
(Section~\ref{hunkdetect:sec}). From key idea 1, we first use a SBFL
tool to locate buggy statements with suspiciousness scores. Hunk
detection algorithm uses those statements to derive the buggy
hunks that need to be fixed~together.





{\bf Multi-Statement Expansion} (Section~\ref{expansion:sec}).
Because SBFL might return one statement for a hunk, we aim to expand
to potentially include more consecutive buggy statements. To do so, we
combine RNN~\cite{Cho-2014} and data-flow analysis to detect more
buggy statements.

{\bf Tree-based Code Repair} (Section~\ref{treerepair:sec}).
For the detected buggy statements from multi-statement expansion, we
use key idea 2 to derive fixes to multiple buggy subtrees at once.
For a buggy subtree~$T$, we build the AST of the method as the
context, and use it as the input of the trained context learning model
(CTL) to produce the weight representing the impact of the
context. The buggy subtree $T$ is used as the input of the trained
tree transformation model (TTL) to produce the context-free
fixed subtree $T'$. Finally, that weight is used to adjust $T'$ into
the fixed subtree $T''$ for a candidate patch.  We apply grammatical
rules and program analysis on the current candidate code to produce
the fixed code. We re-rank and validate the fixed code using test
cases in the same manner as in DLFix~\cite{icse20}.

\section{Training Process}

\subsection{Pairing Buggy and Fixed Subtrees}
\label{pair:sec}

The training data contains the pairs of the source code of the methods
before and after the fixes. Note that a fix might involve multiple
methods. Instead of pairing the entire buggy method with the fixed one,
we use a divide-and-conquer strategy to help the model to better learn
the fixing transformations in the proper contexts.  First, we use the
CPatMiner tool~\cite{icse19-cpatminer} to derive the fixing changes.

If a subtree corresponds to a statement, we call it {\em statement
subtree}. From the result of CPatMiner, we use the following rules to
pair the buggy subtrees with the corresponding fixed subtrees:

1. A buggy subtree ($S$-subtree) is a subtree with
\code{update} or \code{delete}.

2. If a $S$-subtree is \code{deleted}, we pair it with an empty tree.

3. If a buggy $S$-subtree is marked as \code{update}, (i.e, it is
{\em updated} or its children node(s) could be {\em inserted, deleted} or {\em
  updated}), we paired this buggy $S$-subtree with its corresponding
fixed $S$-subtree.

4. If a $S$-subtree is \code{inserted} and its parent node
is another $S$-subtree, we pair it with that parent $S$-subtree.  If
the parent node is not an $S$-subtree, we pair an empty tree to the
corresponding inserted $S$-subtree.




\subsection{Context Building}
\label{fixing-context}

\begin{figure}[t]
	\centering
        \includegraphics[width=3.4in]{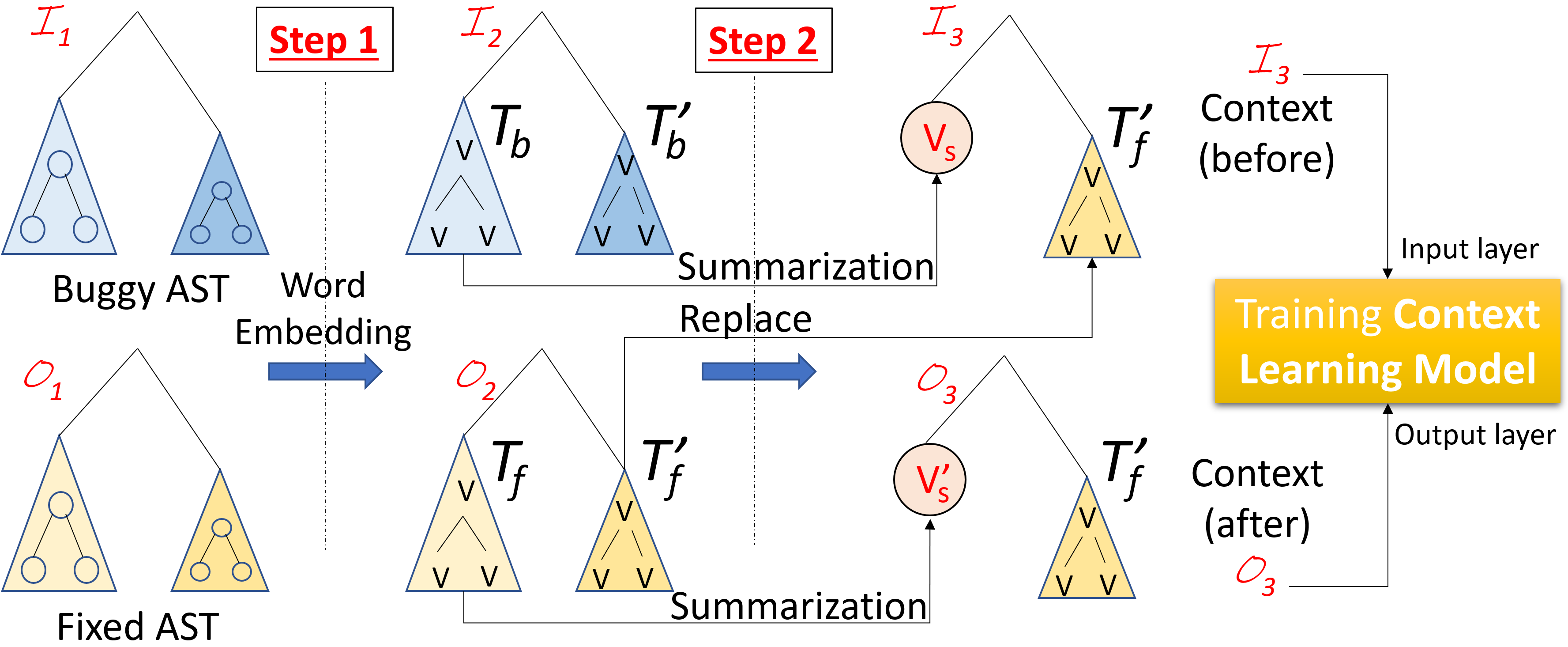} 
	\vspace{-18pt}
	\caption{Context Building to Train Context Learning Model}
	\label{Fig:example_context}
	\vspace{-5pt}
\end{figure}

Figure~\ref{Fig:example_context} illustrates our context building process.
For each pair of the buggy AST $I_1$ and fixed AST $O_1$
(Section~\ref{pair:sec}), we perform alpha-renaming on the variables.
In Step 1, we encode each AST node with the vector using the word
embedding model GloVe~\cite{glove2014} (which captures well code
structure) by considering a statement node as a sentence and each code
token as a word. We use those vectors to label the AST nodes in $I_1$
and $O_1$. The ASTs after this step are the vectorized ASTs $I_2$ and
$O_2$, before and after the fix.

In Step 2, we process each pair of the buggy $S$-subtree $T_b$ in $I_2$
and the corresponding fixed $S$-subtree $T_f$ in $O_2$. First, we
perform node summarization on $T_b$ and $T_f$ by using
TreeCaps~\cite{treecaps} to capture the tree structures of $T_b$ and
$T_f$ into $V_s$ and $V'_s$, respectively. Second, for each of the
other buggy $S$-subtrees, e.g., $T'_b$, and their corresponding fixed
$S$-subtrees, e.g., $T'_f$, we process as follows. Because $T'_f$ is
the fixed version of $T'_b$, we replace $T'_b$ with $T'_f$ in the
building of the resulting context $I_3$ before fixing (key idea 3).
That is, we replace each of the other buggy subtrees with its fixed
version. However, to build the resulting context $O_3$ after fixing,
we keep $T'_f$ because it is the fixed subtree, thus, providing the
correct context.

The resulting AST, $I_3$, is used as the before-the-fix context for
the buggy $S$-subtree $T_b$ and used at {\em the input layer} of the
encoder in the context learning model (CTL). The resulting AST, $O_3$,
is used as the after-the-fix context for $T_f$ and used at {\em the
  output layer} of the decoder in CTL
(Figure~\ref{Fig:example_context}). Finally, the vectors $V_s$ and
$V'_s$ will be used as the weighting inputs for tree
transformation learning later.



%

\subsection{{\bf Context Learning via Tree-based LSTM with Attention Layer and Cycle Training}}
\label{model:sec}

\begin{figure}[t]
	\centering
	\includegraphics[width=3in]{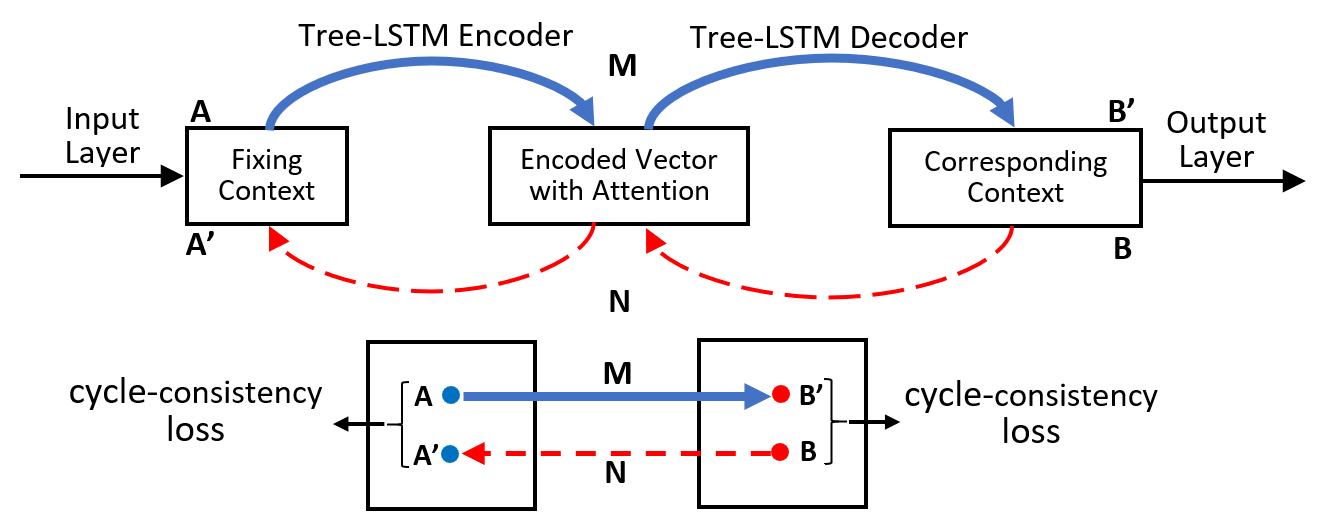}
	\vspace{-8pt}
	\caption{Cycle Training in Attention-based Tree-based LSTM}
	\label{Fig:model}
	\vspace{-4pt}
\end{figure}


For context learning and tree transformation learning, we enhance the
two-layer, tree-based, LSTM models in DLFix~\cite{icse20} with
attention layer and cycle training.
We have added an attention layer into that model, which now
has 3 layers: encoder layer, decoder layer and attention layer
(Figure~\ref{Fig:model}). For the encoder and decoder, to learn the
fixing context expressed in ASTs, we use Child-Sum Tree-based LSTM
\cite{Tai-2015}. Unlike the regular LSTM that loops for each time
step, this model loops for each subtree to capture structures.

We also use {\em cycle training}~\cite{cycle_training} for further
improvement. Cycle training aims to help a model learn better the
mapping between the input and output by continuing to train and
re-train to emphasize~on the mapping between them. This is helpful in
the situations in which a buggy code can be fixed in multiple ways
into different fixed code, or multiple buggy code can be fixed into
one fixed code. This makes the regular tree-based LSTM less
accurate.  With cycle training, the pair of an input and the most
likely output is emphasized to reduce the noise of such one-to-many or
many-to-one~relations.


Cycle training occurs between encoder and decoder. We use the forward
mapping $M$: $A \rightarrow B$ to denote the process of $encoder$
$\rightarrow$ $attention$ $\rightarrow$ $decoder$, and the backward
mapping $N$: $B \rightarrow A$ to denote the process of $decoder$
$\rightarrow$ $attention$ $\rightarrow$ $encoder$
(Figure~\ref{Fig:model}). We apply the adversarial losses for both
$M$ and $N$ to get the two loss functions $L_{run}(M, D_B, A, B)$ and
$L_{run}(N,$ $D_A, B, A)$.  The difference between $N(M(A))$ and $A$,
and that between $M(N(B))$ and $B$ are used to generate
cycle-consistency loss $L_{cyc}(M, N)$ for $M$ and $N$ to ensure the
learned mapping functions are cycle-consistent.
Mathematically, we have two loss functions $L_{run}(M, D_B, A, B)$ and
$L_{run}(N,$ $D_A, B, A)$. With the incentive cycle consistency loss
$L_{cyc}(M,$ $N)$, the overall loss function is computed as follows:
\begin{equation}\label{eq:13}
\begin{split}
L_{cyc}(M, N) &= E_{b\sim p_{data}(b)}[||N(M(a))-a||_1]\\
 &+ E_{a\sim p_{data}(a)}[||M(N(b))-b||_1]
\end{split}
\end{equation}
\begin{equation}\label{eq:14}
\begin{split}
L(M,N,D_{a},D_{b}) &= L_{run}(M, D_B, A, B)
+ L_{run}(N, D_A, B, A)\\
+ \alpha L_{cyc}(M, N)
\end{split}
\end{equation}
Where $L(M,N,D_{a},D_{b})$ is the loss function for the entire cycle
training; $M$ and $N$ are the mapping functions to map $A$ to $B$ and $B$ to
$A$; $D_A$ is aimed to distinguish between the predicted result
$N(M(A))$ and the real result $A$; $D_B$ is aimed to distinguish
between the predicted result $M(N(B))$ and the real result $B$;
$L_{run}$ is the cycle consistency loss function for the running
function $M,N$; $L_{cyc}$ is the incentivized cycle consistency loss; and
$\alpha$ is the parameter to control the relative importance of the
two objectives.

\subsection{Tree Transformation Learning}
\label{codetrans:sec}

\begin{figure}[t]
	\centering
        \includegraphics[width=3.4in]{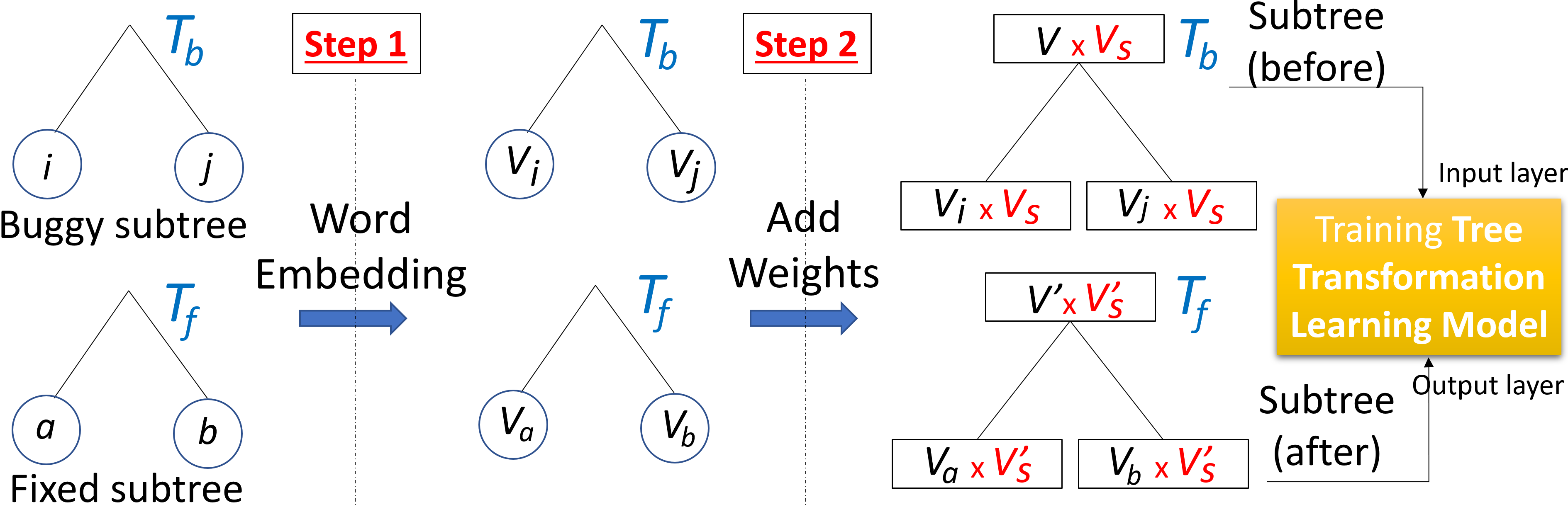}
	\vspace{-15pt}
	\caption{Tree Transformation Learning ($V_S$, $V'_S$ in
          Figure~\ref{Fig:example_context})}
	\label{Fig:example_trans}
\end{figure}

Figure~\ref{Fig:example_trans} illustrates the tree transformation learning process.
We use the same tree-based LSTM model with attention layer and cycle
training as in Section~\ref{model:sec} to learn the code
transformation for each buggy $S$-subtree $T_b$. In Step 1, we build
the word embeddings for all the code tokens as in
Section~\ref{fixing-context}.
Each AST node in the buggy $S$-subtree ($T_b$) and the fixed one
($T_f$) is labeled with its vector representation
(Figure~\ref{Fig:example_trans}). Next, we use the summarized vectors
$V_s$ and $V'_s$ computed from context learning in
Figure~\ref{Fig:example_context} as the weights and perform
cross-product for each vector of the node in the buggy $S$-subtree
$T_b$ and for each one in the fixed $S$-subtree $T_f$, respectively.
The two resulting subtrees after cross-product are used at the input
and output layers of the tree-based LSTM model for tree transformation
learning. We use cross-product because we aim to have a vector as the
label for a node and use it as a weight representing the context to
learn code transformations for bug fixes.

\section{Fixing Process}


\subsection{Fixing-together Hunk Detection Algorithm}
\label{hunkdetect:sec}


The first step of fixing multi-hunk, multi-statement bugs is for
our FL method to 
{\em detect buggy hunk(s) that are fixed together in the same
  patch}. To do that, we fine-tune Google's pre-trained BERT
model~\cite{devlin2018bert} to learn the {\em fixing-together
  relationships among statements} using BERT's sentence-pair
classification task. Then, we use the fine-tuned BERT model
in an algorithm to detect fixing-together hunks. Let us explain
our hunk detection algorithm in details.


\subsubsection{\bf Fine-tuning BERT to Learn Fixing-together Relationships among Statements}

We first fine-tune BERT to learn if two statements are needed to
be fixed together or not. Let $H$ be a set of the hunks that are fixed
together for a bug. The input for the training process is all the sets
$H$s for all the bugs in the training set.


\subsubsection*{\underline{Step 1}}
For a pair of hunks $H_i$ and $H_j$ in $H$, we take every pair of
statements $S_k$ and $S_l$, one from each hunk, and build the vectors
with BERT. We consider the pair of statements ($S_k$, $S_l$) as being
fixed-together in the same patch to fine-tune BERT.


\subsubsection*{\underline{Step 2}}
Step 1 is repeated for all the pairs of the
statements ($S_k$, $S_l$)s in all the pairs $H_i$ and $H_j$ in $H$.
We also repeat Step 1 for all $H$s. We use them to fine-tune the BERT
model to learn the fixing-together relationships among any two
statements in all pairs of hunks.


\subsubsection{\bf Using Fine-tuned BERT for Hunk Detection}
After obtaining the fine-tuned BERT, we use it in determining whether
the hunks of code need to be fixed together or not. The input of of
this procedure is the fine-tuned BERT model, buggy code $P$, and test
cases. The output is the groups of hunks that need to be fixed
together. The process is conducted in the following steps.


\subsubsection*{\underline{Step 1}}

We use a spectrum-based FL tool (in our
experiment, we used Ochiai~\cite{abreu2006evaluation}) to run on the
given source code and test cases. It returns the list of buggy
statements and suspiciousness scores.


\subsubsection*{\underline{Step 2}}

The consecutive statements within a method
returned by the FL tool are grouped together to form the hunks $H_1$,
$H_2$, ..., $H_m$.

\subsubsection*{\underline{Step 3}}
To decide if a pair of hunks ($H_i$,$H_j$) needs to be fixed together,
we use the BERT model that was fine-tuned. Specifically, for every
pair of statements ($S_k$,$S_l$), one from each hunk ($H_i$,$H_j$), we
use the fine-tuned BERT to measure the fixing-together relationship
score for ($S_k$,$S_l$). The fixing-together score between $H_i$ and
$H_j$ is the average of the scores of all the pairs of statements
within $H_i$ and $H_j$, respectively.
If the average score for all the statement pairs is higher than a
threshold, we consider ($H_i$,$H_j$) as needed to be fixed
together. From the pairs of the detected hunks, we build the groups of
the fixing-together hunks. The group of hunks that has any statement with
the highest Ochiai's suspiciousness score will be ranked and fixed
first. The rationale is that such a group contains the most suspicious
statement, thus, should be fixed first.

\subsection{Multiple-Statement Expansion Algorithm}
\label{expansion:sec}



A detected buggy hunk from the algorithm in
Section~\ref{hunkdetect:sec} might contain only one statement since
each of those suspicious statements is originally derived by a SBFL
tool, which does not focus on detecting consecutive buggy statements
in a hunk. Thus, in this step, we take the result from the hunk
detection algorithm, and expand it to include potentially more
statements in a hunk.


\subsubsection*{\underline{Key Idea}}
Our idea is to combine deep learning with data flow analysis. We first
train an RNN model with GRU cells~\cite{Cho-2014} (will be explained
in Section~\ref{rnn:sec}) to learn to decide whether a statement is
buggy or not. We collect the training data for that model from the
real buggy statements. We then use data-flow analysis to adjust the
result.  Specifically, if a statement is labeled as buggy by the RNN
model, no adjustment is needed. However, even when the RNN model
decides a given statement $s$ as {\em non-buggy}, and if $s$ has a
data dependency with a buggy statement, we still mark $s$ as {\em
  buggy}.





\subsubsection{{\bf Expansion Algorithm}}

The input of Multi-Statement Expan\-sion algorithm is the buggy
statement \code{buggyS}, i.e., the seed statement of a hunk. The
output is a buggy hunk of consecutive statements.

\begin{algorithm}[t]
\caption{Multiple-Statement Expansion Algorithm}
\label{algo}
\scriptsize
\begin{algorithmic}[1]
\Function {MultiStatementsExpansion}{$buggyS$}
\State $candStmts = Expand2NCandidatesList(buggyS)$
\State $predResult = RNNClassifier(candStmts)$
\State $expandResult = DataDepAnalysis(candStmts,predResult)$
\State \textbf{return} $expandResult$
\EndFunction






\Function {DataDepAnalysis} {$buggyS$,$candStmts$, $predResult$}
	\State $buggyHunk = GetCenterBuggyHunk(predResult)$
	\State $DDExpandHunk(buggyS,buggyHunk, TopHalf(candStmts))$
	\State $DDExpandHunk(buggyS,buggyHunk, BotHalf(candStmts))$
	
	\State \textbf{return} $buggyBlock$
\EndFunction

\Function {DDExpandHunk} {$buggyS$, $buggyHunk$, $candStmts$}
\For {\textbf{each} $(stmt \in candStmts)$ $\textbf{\&}$ $(stmt \notin buggyHunk)$}
\If {$HasDataDep(stmt, buggyS)$}
	\State $buggyHunk = buggyHunk \cup stmt$
\Else { \textbf{break}}
\EndIf
\EndFor
	\State \textbf{return} $buggyHunk$
\EndFunction
\end{algorithmic}
\end{algorithm}


First, it produces a candidate list of buggy statements by including
$N$ statements before and $N$ statements after \code{buggyS} ({\em
  Expand2N\-CandidatesList} at line 2). In the current implementation,
$N$=5.  Then, it uses the RNN model to act as a classifier to predict
whether~each statement (except \code{buggyS}) in the candidate
list is buggy or not ({\em RNNClassifier} at line 3). To train that
RNN model, we use the buggy statements in the buggy hunks in the
training data (see Section~\ref{rnn:sec}). TreeCaps~\cite{treecaps}
is used to encode the statements.
%

\begin{figure}[t]
	\centering
	\includegraphics[width=3in]{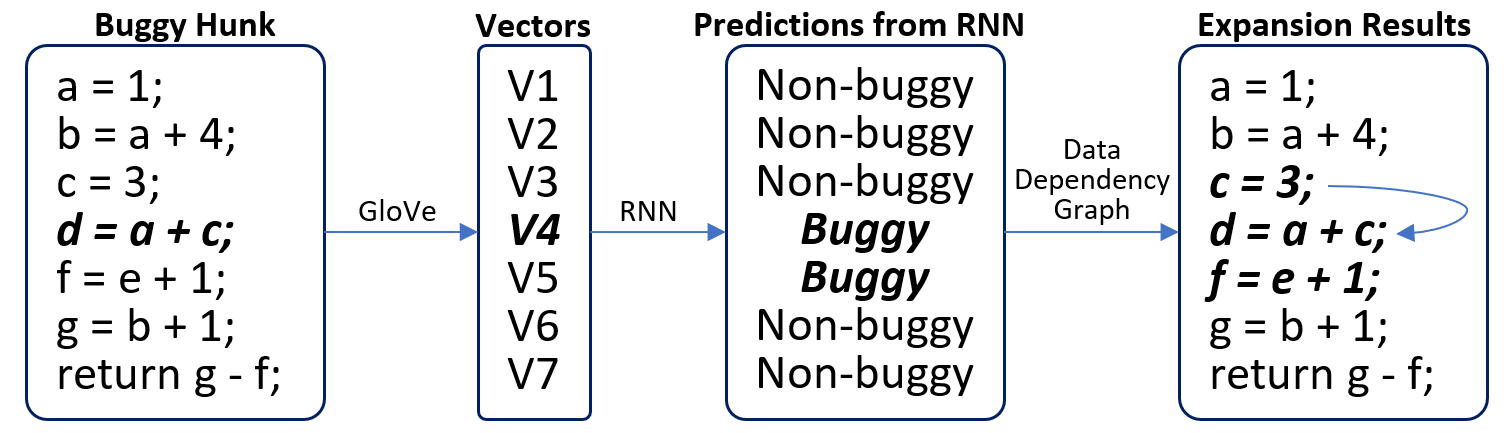}
        \vspace{-0.12in}
	\caption{Multiple-Statement Expansion Example}
	\label{Fig:expansion}
\end{figure}

In {\em DataDepAnalysis} (line 4), to adjust the results from the RNN
model, we obtain {\em buggyHunk} surrounding the buggy statement
\code{buggyS}, consisting of the statements before and after
\code{buggyS}, that~were predicted as buggy by the RNN model (line
7). We then examine statement-by-statement in the upward direction
from the center buggy statement in the candidate list
(line 8, via {\em TopHalf}) and in the downward direction
(line 9, via {\em BotHalf}). In {\em DDExpandHunk},~we continue
to expand (upward or downward) the current buggy hunk {\em buggyHunk}
to include a statement that is deemed as buggy by~the RNN model
or has a data dependency with the center buggy statement {\em buggyS}
(lines 13--14). We stop the process (upward or downward), if we
encounter a non-buggy statement without data dependency with {\em buggyS} or we
exhaust the list (line 15). Finally, the buggy hunk
containing consecutive buggy statements is returned.

In Figure~\ref{Fig:expansion},
the SBFL tool returns the buggy statement at line 4. All statements
are encoded into the sets of vectors via GloVe~\cite{glove2014} and
classified by the RNN model. We expand from the statement at line 4
upward to include line 3 (even though the RNN model predicted it as
non-buggy), since line 3 has a data dependency with the buggy
statement at line 4 via the variable \code{c}. We include line 5
because the RNN model predicts the line 5 as buggy.
At this time, we stop the upward and downward directions because we
encounter the non-buggy statements at lines 2 and 6 that do not
have data dependency with line 4. That is, lines 1--2 and 6--7 are
excluded. The final result includes the statements at lines 3--5 as
the buggy~hunk.




\subsubsection{\bf Buggy Statement Prediction with RNN}
\label{rnn:sec}

We present how to use an GRU-based RNN model~\cite{Cho-2014} to
predict a buggy statement.


\subsubsection*{Training} To train the RNN model, we use the buggy/non-buggy statements in all
the hunks in the training dataset. We use GloVe~\cite{glove2014} to
encode each token in a statement so that a statement is represented by
a sequence of token vectors. We use the neural architecture of the
GRU-based RNN model~\cite{Cho-2014} to consume the GloVe vectors of
statements associated with the buggy/non-buggy labels.

The RNN model operates in the time steps. At the time step $k$
($k$>=1), at the input layer, GRU consumes the GloVe vectors of the
$k^{th}$ statement $S_k$. At the output layer, $S_k$ is labeled as 1
if it is a buggy statement and as 0 otherwise. In addition to the
input at the time step $k+1$, we feed the output of the time step $k$
to the GRU.


\subsubsection*{Prediction}
The trained GRU-based RNN model is used in Expansion
Algorithm
at line 3 to predict if a statement in the hunk is buggy or not. The
model takes a statement in the form of GloVe token vectors. It takes
the vectors of all the statements in a hunk and labels them as buggy
or non-buggy in multiple-time-step manner.


\subsection{Tree-based Code Repair}
\label{treerepair:sec}

Figure~\ref{Fig:repair} illustrates this process. After deriving the
buggy hunk(s) in the method(s), {\tool} performs code repairing for all
buggy statements in all the hunks at once using the trained LSTM models. The
tree-based code repair is conducted in the following steps:




\subsubsection*{\underline{Step 1.} Identifying buggy S-subtrees}

For each hunk, we parse the code into AST, and identify the buggy
$S$-subtrees corresponding to the derived buggy statements.  In
Figure~\ref{Fig:repair}, the $S$-subtrees $T_1$ and $T_2$ are
identified as buggy. If a buggy $S$-subtree is part of another larger
buggy $S$-subtree, we just need to perform fixing on the larger
$S$-subtree since that fix also fixes the smaller $S$-subtree.



\subsubsection*{\underline{Step 2.} Embedding and Summarization}
\begin{figure}[t]
	\centering
        \includegraphics[width=3.3in]{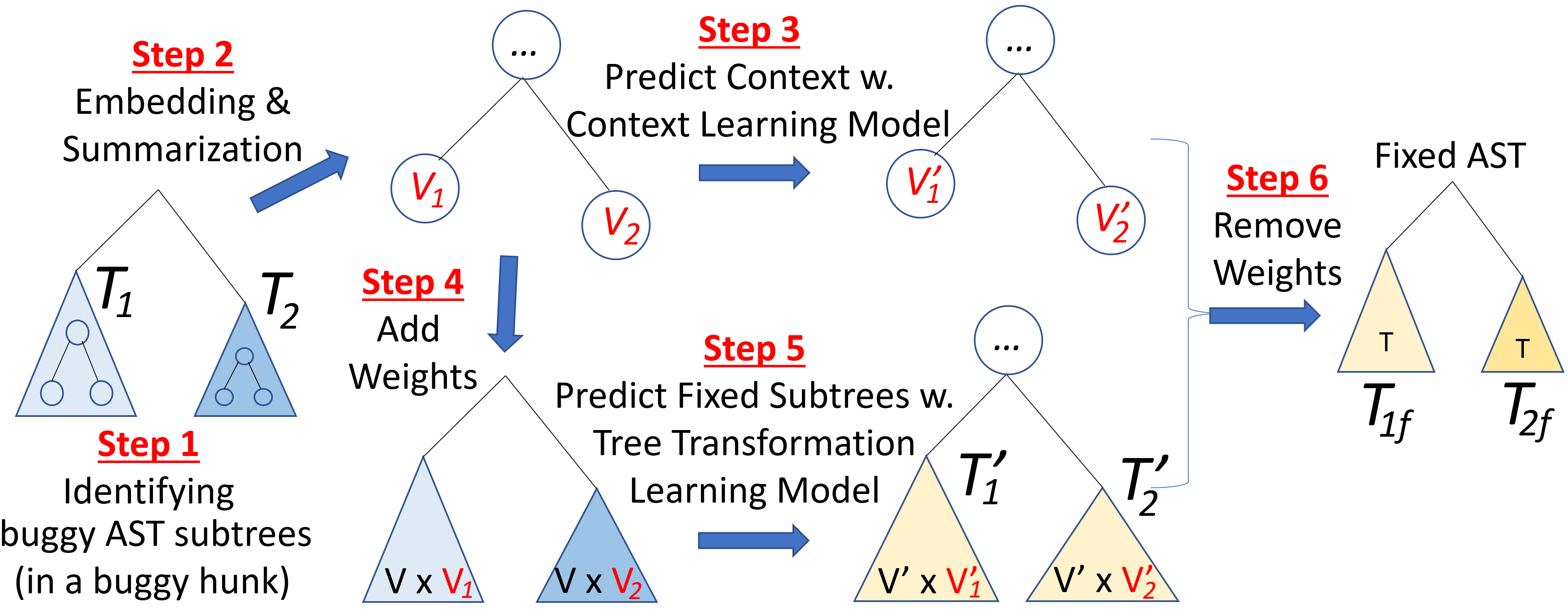}
        \vspace{-6pt}
	\caption{Tree-based Code Repair}
	\label{Fig:repair}
\end{figure}

We perform word embedding using GloVe~\cite{glove2014} and tree
summarization using TreeCaps~\cite{treecaps} on all the buggy subtrees
to obtain the contexts.  For example, in Figure~\ref{Fig:repair},
$T_1$ and $T_2$ are summarized into two vectors $V_1$ and $V_2$.


\subsubsection*{\underline{Step 3.} Predict Context}
We use the trained context learning model (CTL) to run on the context
with the AST nodes including also the summarized nodes to predict the
context. In the resulting AST, the structure is the same as the AST
for the input context, except that the summarized nodes become the new
ones. For example, $V_1$ and $V_2$ in the context becomes $V'_1$ and
$V'_2$ after Step~3.

%


\subsubsection*{\underline{Step 4.} Adding Weights}
The weights $V_1$ and $V_2$ from Step 2 are used in a product with
the vectors in the buggy subtrees $T_1$ and $T_2$. Each node in $T_1$
and $T_2$ is represented by a multiplication vector between the
original vector of the node and the weight vector $V'_1$ or $V'_2$.


\subsubsection*{\underline{Step 5.} Predict Transformations}
We use the trained tree transformation learning model to predict
the subtrees $T'_1$ and $T'_2$ of the~fix.


\subsubsection*{\underline{Step 6.} Removing Weights}
We remove the weight from Step 4 to obtain the candidate fixed subtree
for a buggy one. For example,~we remove $V'_1$ and
$V'_2$ to obtain the candidate fixed subtrees $T_{1f}$ and~$T_{2f}$.
However, because we know the cross product and a vector, we can get
the unlimited number of solutions. Thus, to produce a
single solution $T_{1f}$ and $T_{2f}$, for each node in $T'_1$ and $T'_2$, we assume that
$V'_1$ and $V'_2$ are vertical with the node vector $V_{n1}$ in $T'_1$
and $V_{n2}$ in $T'_2$. Then, we can get the unweighted node vector
$V'_{n1}$ in $T_{1f}$ as follows:
\begin{equation}\label{eq:15}
\begin{split}
V'_{n1} = \frac{V'_1 \times V_{n1}}{V'_1 \dot V'_1}
\end{split}
\end{equation}
After having the vector for each node in a fixed S-subtree $T_{1f}$, we
generated candidate patches based on word embedding. For each node, we
calculated the cosine similarity score between its vector $V'_{n1}$
and each vector in the vector list for all tokens. To generate
candidate patches, we select the token $t$ in the list. Token $t$ has
its similarity score $Score_t$ for one node in the fixed
$S$-subtree. By adding all $Score_t$ for all the tokens, we have the
total score, $Score_{sum}$, for a candidate. We select the top-5
candidates for each node to generate the candidates and sort them
based on $Score_{sum}$.

\subsection{Post-processing}

A naive approach would face combinatorial explosion in forming the
candidate fix(es) because for each node in the sub-tree, we maintain
top-5 candidates. However, when we combine the candidates for all the
nodes in the fixed sub-tree, many candidates are not valid for the
current method in the project. Therefore, when we form a candidate by
combining all the candidates for the nodes, we apply a set of filters
to verify the program semantics in the same manner as in
DLFix~\cite{icse20}. This allows us to eliminate invalid candidates
immediately. Specifically, we use the alpha-renaming filter to change
the names back to the normal Java code using a dictionary containing
all the valid names in the scope, the syntax-checking filter to remove
the candidates with syntax errors, and the name validation filter to
check the validity of the variables, methods, and classes. Moreover,
for further improvement, we use beam search to maintain only the
top-ranked candidate fixes. Thus, we do not exhaust all compositions
in forming the statements. This helps maintain a manageable number of
candidates.

After applying all the filters, we also used DLFix~\cite{icse20}'s
re-ranking scheme on the candidate patches.
%
We then used test cases to conduct patch validation on those
candidates. We verify each patch from the top to the bottom until a
correct patch is identified and the patch validation ends.
If all candidates for fixing a location cannot pass all the
test cases, we select the next location to repeat the process.

\section{Empirical Evaluation}
\label{eval:sec}


\subsection{Research Questions}

To evaluate {\tool}, we seek to answer the following questions:



\noindent\textbf{RQ1. Comparative Study with Deep Learning-based APR models on
Defects4J benchmark.}  How well does {\tool} perform in comparison with
existing {\em DL-based} APR models on {\em Defects4J}?


\noindent\textbf{RQ2. Comparative Study with Deep Learning-based APR
  models on Large Bug Datasets.}  How well does {\tool} perform in
comparison with {\em DL-based} APR models on {\em large-scale bug
  datasets}?

\noindent\textbf{RQ3. Comparative Study with Pattern-based APR approaches on Defects4J.} 
How well does {\tool} perform in comparison with the state-of-the-art, {\em pattern-based} APR approaches?


\noindent\textbf{RQ4. Sensitivity Analysis of {\tool}.} 
How do various factors affect the overall performance of {\tool} in APR?

\noindent\textbf{RQ5. Time Complexity and Model's Training Parameters.} 
What is time complexity and the numbers of training parameters?


\subsection{Data Collection}
\label{dataset}


We have conducted our empirical evaluation on three datasets:

1) {\it Defects4J} v1.2.0~\cite{defects4j} with 395 bugs with test
cases;

2) {\it BigFix}~\cite{icse20} with +26k bugs in +1.8
million buggy methods;

3) {\it CPatMiner}~\cite{icse19-cpatminer} with +44k bugs
from 5,832 Java projects.

All experiments were conducted on a workstation with a 8-core Intel
CPU and a single GTX Titan~GPU.





\subsection{Experimental Methodology}


\subsubsection{\bf RQ1. Comparison with DL-based APR on Defects4J}

{\em \underline{Comparative Baselines.}}
We compare {\tool} with five state-of-the-art DL-based APR
models: \textbf{DLFix}~\cite{icse20}, \textbf{CoCoNuT}~\cite{lutellier2020coconut}, \textbf{SequenceR}~\cite{chen2018sequencer}, \textbf{Tufano19}~\cite{tufano2019learning}, \textbf{CODIT}~\cite{chakrabortycodit},
and {\bf CURE}~\cite{cure-icse21}.


\vspace{2pt}
{\em \underline{Procedure and Settings.}}  
We replicated all DL-based APRs except CURE, which is unavailable.
We re-implemented CURE following the details in their paper.
We trained all DL approaches on the bugs and fixes in CPatMiner
dataset and tested them on all 395 bugs in Defects4J (no overlap
between the two datasets). All DL approaches were applied with the
same fault localization tool, Ochiai~\cite{abreu2006evaluation}, and
patch validation with the test cases in Defects4J.
%
Following prior experiments~\cite{Simfix,icse20}, we set a 5-hour
running-time limit for a tool for patch generation and validation.


We tuned {\tool} with the following key hyper-parameters using the
beam-search: (1) BERT for hunk detection: epoch size (e-size) $(2, 3,
4, 5)$, batch size (b-size) $(8, 16, 32, 64)$, and learning rate
(l-rate) $(3e^{-4}, 1e^{-4}, 5e^{-5}, 3e^{-5}, 1e^{-5})$; (2) LSTM for
Multi-Statement Expansion and code repair: e-size $(100, 150, 200,
250)$, b-size $(32, 64, 128, 256)$, and l-rate $(0.0001, 0.0005,
0.001, 0.003, 0.005)$; (3) GloVe for representation vectors: vector
size (v-size) $(100, 150, 200, 250)$, l-rate $(0.001, 0.003, 0.005,
0.01)$, b-size $(32, 64, 128, 256)$, and e-size $(100, 150$, $200,
250)$. The other default parameters were used.

The best setting for DEAR is (1) e-size=4, b-size=32, l-rate=$1e^{-4}$
for BERT; (2) e-size=200, l-rate=0.003, b-size=128 for LSTM; (3)
v-size=200, l-rate=0.001, b-size=64, e-size=200 for GloVe.  For other
models, we tuned with the parameters in their papers, e.g., the vector
length of word2vec, learning rate, and epoch size to find the best
parameters for each dataset.
We tuned all approaches with the aforementioned parameters on the
same {\it CPatMiner} dataset to obtain the best performance.  Once we
obtained the best parameters for each model, we used them for
later experiments.


\vspace{2pt}
{\em \underline{Quantitative Analysis.}}  We report the numbers of bugs
that a model can auto-fix
for the following bug-location types:
	
	{\em \bf Type-1. One-Hunk, One-Statement}: A bug with the fix involving only one hunk with one single statement.
	
	{\em \bf Type-2. One-Hunk, Multi-Statements}: A bug with the fix involving only one hunk with multiple statements.
	
	{\em \bf Type-3. Multi-Hunks, One-Statement}: A bug with the fix involving multiple hunks; each hunk with one fixed statement.
	
	{\em \bf Type-4. Multi-Hunks, Multi-Statements}: A bug with the fix involving multi-hunks; each hunk has multiple statements.
	
	{\em \bf Type-5. Multi-Hunks, Mix-Statements}: A bug with the fix involving multiple hunks, and some hunks
	have one statement and other hunks have multiple statements.


\vspace{1pt}
{\em \underline{Evaluation Metrics.}}
We report the {\em number of bugs} that can be correctly fixed
and the number of plausible patches (i.e., passing all test cases, but
not the actual fixes) using the {\em top candidate patches}.



\subsubsection{{\bf RQ2. Comparison with DL-based APR on Large Datasets}}


{\em \underline{Comparative Baselines.}} We compare {\tool} with the same
baselines as in  RQ1 on two large datasets: BigFix and CPatMiner.
%



\vspace{1pt}
{\em \underline{Procedure and Settings.}}  First, we evaluated all
DL-based APR~models on BigFix and CPatMiner. Following DLFix and
Sequencer, we randomly split data into 80\%/10\%/10\% for
training, tuning, and testing.
Second, we have cross-dataset evaluation: training DL-based approaches
on CPatMiner and testing on BigFix, and vice versa.
Unlike Defects4J, BigFix and CPatMiner datasets do not have test
cases. Without test cases, we cannot use fault localization and patch
validation for all DL approaches. Thus, we fed the actual bug
locations into the DL models, including locations on buggy hunks and
statements. The DL-based baselines do not distinguish hunks, instead
{\em process each buggy statement at a time}. We use developers'
actual fixes as the ground truth to evaluate the DL-based approaches.




\vspace{1pt}
{\em \underline{Evaluation Metrics.}}  We use the {\em top-$K$
metric}, defined as the ratio between the number of times that a
correct patch is in a ranked list of the top $K$ candidates over the total
number of bugs.

\subsubsection{{\bf RQ3. Comparison with Pattern-based APR on Defects4J}}

{\em \underline{Comparative Baselines.}} We compare {\tool} with the
	 state-of-the-art, pattern-based APR tools on
	 Defects4J: \textbf{Elixir}~\cite{saha2017elixir}, \textbf{ssFix}~\cite{xin2017leveraging}, \textbf{CapGen}~\cite{wen2018context}, \textbf{FixMiner}~\cite{koyuncu2018fixminer}, \textbf{Avatar}~\cite{liu2019avatar}, \textbf{Hercules}~\cite{hercules-icse19}, \textbf{SimFix}~\cite{Simfix},
	 and \textbf{Tbar}~\cite{tbar-issta19}. We were able to
	 replicate the following pattern-based baselines: {\bf Elixir,
	 ssFix, FixMiner, SimFix, TBar} under the same computing
	 environments. We set the time limit to 5 hours for the
	 tools. For the other baselines, due to unavailable code, we
	 use the results reported in their papers as they were run on
	 the same dataset. We used the same setting and evaluation metric.

\subsubsection{{\bf RQ4. Sensitivity Analysis}}
We evaluate the impacts of different factors on {\tool}'s
performance. We consider the following: (1) hunk detection
(Hunk); (2) multi-statement expansion (Expansion); (3) multi-statement
tree model and cycle training; and (4) data splitting scheme. We use
the left-one-out strategy for each factor.  We evaluate the first
three factors on Defects4J and the last one on CPatMiner since we need
a larger dataset for various splitting.






\subsubsection{{\bf Time Complexity and Numbers of Parameters in Model Training}}

We measure the training and fixing time for a model and its number of parameters for model training on the datasets.


\section{Empirical Results}
\label{result:sec}


\subsection{\bf  RQ1. Comparison Results with DL-based APR Models on Defects4J}
\label{rq1:sec}

\begin{table}
	\caption{RQ1. Comparison with DL-based APR Models on Defects4J \underline{with} Fault Localization}
	\vspace{-10pt}
	\begin{center}
        \small
		\renewcommand{\arraystretch}{1} 
		\begin{tabular}{l|p{0.6cm}<{\centering}p{0.6cm}<{\centering}p{0.6cm}<{\centering}p{0.6cm}<{\centering}p{0.7cm}<{\centering}p{0.6cm}<{\centering}|p{0.4cm}<{\centering}}
			\hline
			Projects&  Chart & Closure & Lang & Math & Mockito & Time & Total\\
			\hline
		    Sequencer       & 3/3   & 4/5     & 2/2  & 6/9  & 0/0     & 0/0  & 15/19 \\
			CODIT   		& 1/2  &  2/5   & 0/0 & 3/5 &  0/0  & 0/0 & 6/12  \\
			Tufano19 		& 3/4  &  3/5   & 1/1 & 6/8 &  0/0  & 0/0 & 14/18  \\
			DLFix     	    & 5/12  & 6/10    & 5/12 & 12/18& 1/1     & 1/2  & 30/55\\
			CoCoNuT         & 6/11 & 6/9 & 5/13 & 13/21 & 2/2  & 1/1 & 33/57 \\
            CURE      &  6/13   &  6/10    &  5/14   &   16/23   &    2/2    & 1/2 & 36/71 \\
			\hline
			{\bf \tool}   	& 8/16  & 7/11    & 8/15 & 20/33 & 1/2     & 3/6  & \textbf{47/91}\\
			\hline
		\end{tabular}
		X/Y:  are the numbers of correct and plausible patches, respectively.
		\label{RQ1}
	\end{center}
\vspace{-2pt} 
\end{table}

\begin{table}
	\caption{RQ1. Detailed Comparison with DL-based APR Models on Defects4J \underline{with} Fault Localization}
	\vspace{-10pt}
	\begin{center}
		\small
		\renewcommand{\arraystretch}{1} 
		\begin{tabular}{p{3.9cm}<{\centering}|p{0.6cm}<{\centering}|p{1cm}<{\centering}|p{0.6cm}<{\centering}<{\centering}|p{0.6cm}}
			\hline
			Bug Types & DLFix& CoCoNuT  & CURE & {\tool}\\\hline
			
			Type 1. One-Hunk One-Stmt & 30 & 33& 36 & 29\\
			Type 2. One-Hunk Multi-Stmts  & 0 & 0& 0 & 4\\ 
			Type 3. Multi-Hunks One-Stmt  & 0 & 0& 0 & 11\\
			Type 4. Multi-Hunks Multi-Stmts  & 0 &0& 0 & 1\\
			Type 5. Multi-Hunks Mix-Stmts  & 0 & 0& 0 & 2\\\hline
			Total & 30 & 33 & {\bf 36} & {\bf 47} \\
			\hline
		\end{tabular}
		\label{RQ1_1}
	\end{center}
\vspace{-3pt}
\end{table}

\subsubsection{{\bf With Fault Localization}}

We first evaluate the APR models when using with the fault
localization tool Ochiai~\cite{abreu2006evaluation}.
Tables~\ref{RQ1} and ~\ref{RQ1_1} show the comparison results
among {\tool} and the baseline models.

As seen in Table~\ref{RQ1}, {\tool} can auto-fix the most number of
bugs (47) and generate the most number of plausible patches (91) that
pass all test cases on Defects4J. Particularly, {\tool} can auto-fix
32, 41, 33, 17, 14, and 11 more bugs than Sequencer, CODIT,
Tufano19, DLFix, CoCoNuT, and CURE, respectively (i.e., 213\%, 683\%,
236\%, 57\%, 42\%, and 31\% relative improvements).  Compared
with
those tools in that order on Defects4J, {\tool} can auto-fix 35, 34,
41, 18, 31, and 18 bugs that those tools missed, respectively. Via the
overlapping analysis between the result of {\tool} and those of the
baselines combined, {\tool} can fix 18 unique bugs that they missed.







Table~\ref{RQ1_1} shows the comparison between {\tool} and
the top DL-based baselines (DLFix, CoCoNuT, CURE) w.r.t. different
bug types.

For \underline{single-hunk bugs} (Types 1-2), {\tool} fixes 33 bugs
including 4 unique single hunk bugs that the other tools missed.

For \underline{multi-hunk bugs} (Types 3--5), {\tool} can fix 14 bugs
that cannot be fixed by DLFix, CoCoNuT, and CURE. Existing
DL-based APR models cannot fix those bugs since the mechanism of
fixing one statement at a time does not work on the bugs that require
the fixes with dependent changes to multiple statements at
once. Thus, they do not produce correct patches for those cases.

For \underline{multi-hunk or multi-statement bugs} (Types 2--5), {\tool}
fixes 18 of them (out of 47 fixed bugs, i.e., 38.3\% of total fixed
bugs).




\subsubsection{{\bf Without Fault Localization}}

We also compared {\tool} with other tools in the fixing capabilities
without the impact of a third-party FL tool. All the tools under
comparison (Table~\ref{RQ1_1_1}) were pointed to the correct fixing
locations and performed the fixes. As seen, if the fixing locations
are known, {\tool}'s fixing capability is also higher than those
baselines (53 bugs versus 44, 40, and 48). Importantly, it can fix 20
multi-hunk/multi-statement bugs (37.7\% of a total of 53 fixed bugs),
while CoCoNuT, DLFix, and CURE can fix only 7, 5, and 10 such bugs.



{\tool} is more general than existing DL-based models because it can
support dependent fixes with multi-hunks or
multi-statements. Importantly, {\em it significantly improves these
DL-based models and raises the DL direction to the same level as the
other APR directions (search-based and pattern-based)}, which can
handle multi-statement bugs. Moreover, {\tool} is fully data-driven and
does not require the defined fixing patterns as in the pattern-based APR models.

\begin{table}
	\caption{RQ1. Comparison with DL-based APR Models on Defects4J \underline{without} Fault Localization (i.e., Correct Location)}
	\vspace{-7pt}
	\begin{center}
		\small
		\renewcommand{\arraystretch}{1} 
\begin{tabular}{p{3.9cm}<{\centering}|p{0.6cm}<{\centering}|p{1cm}<{\centering}|p{0.6cm}<{\centering}<{\centering}|p{0.6cm}}
			\hline
			Bug Types & DLFix& CoCoNuT & CURE & {\tool}\\\hline
			
			Type 1. One-Hunk One-Stmt  & 35 & 37 & 38 & 33\\
			Type 2. One-Hunk Multi-Stmts  & 1 &3 & 3 & 4\\ 
			Type 3. Multi-Hunks One-Stmt  & 4 &3 & 6 & 13\\
			Type 4. Multi-Hunks Multi-Stmts  & 0 &0 & 0 & 1\\
			Type 5. Multi-Hunks Mix-Stmts  & 0 & 1 & 1 & 2\\\hline
			Total & 40 & 44 & 48 & {\bf 53} \\
			\hline
		\end{tabular}
		\label{RQ1_1_1}
	\end{center}
	\vspace{-4pt} 
\end{table}




\subsection{RQ2. Comparison Results with DL-based APR Models on Large Datasets}
\label{rq2:sec}


\begin{table}[t]
	\caption{RQ2. Comparison with DL APRs on Large Datasets}
	\vspace{-8pt}
	\begin{center}
        \small
		\renewcommand{\arraystretch}{1}
		\begin{tabular}{l|p{0.7cm}<{\centering}|p{0.7cm}<{\centering}|p{0.7cm}<{\centering}|p{0.7cm}<{\centering}|p{0.7cm}<{\centering}|p{0.7cm}<{\centering}}
			\hline
			 \multirow{2}{*}{} & \multicolumn{3}{c|}{\shortstack{CPatMiner\\(4,415 tested bugs)}} & \multicolumn{3}{c}{\shortstack{BigFix \\(2,594 tested bugs)}} \\
			\cline{2-7}
		Tool/Dataset	& Top-1  & Top-3 & Top-5 & Top-1 & Top-3 & Top-5 \\
			\hline
				Sequencer & 7.8\%  & 8.9\% & 10.3\%& 8.5\% & 9.1\% & 10.8\%\\
                CODIT     &4.5\%   & 7.4\% & 9.2\% &3.9\%  &6.3\%  &9.1\%  \\ 				
				Tufano19    & 8.6\%  & 9.3\% & 11.2\%& 7.7\% & 8.8\% &9.6\% \\
			    DLFix     & 11.4\% & 12.3\%& 13.1\%& 11.2\%& 11.9\%& 12.5\%\\
			    CoCoNuT   & 13.5\% & 14.7\%& 15.3\% & 12.2\% & 13.6\% & 14.3\% \\
                        CURE   & 14.2\%& 15.1\% & 15.5\% & 12.9\% & 14.2\% & 14.1\% \\    
			\hline
			{\tool}   & \textbf{15.1\%} & \textbf{15.6\%}& \textbf{16.8\%}& \textbf{14.1\%}& \textbf{15.4\%}& \textbf{16.3\%}\\			
			\hline
		\end{tabular}
		\label{RQ2}
		\vspace{-2pt}
	\end{center}
\end{table}

\begin{table}[t]
	\caption{RQ2. Comparison with DL APRs on Cross-Datasets}
	\vspace{-8pt}
	\begin{center}
        \small
		\renewcommand{\arraystretch}{1}
		\begin{tabular}{l|p{0.7cm}<{\centering}|p{0.7cm}<{\centering}|p{0.7cm}<{\centering}|p{0.7cm}<{\centering}|p{0.7cm}<{\centering}|p{0.7cm}<{\centering}}
			\hline
			\multirow{2}{*}{Tool/Dataset} & \multicolumn{3}{c|}{CPatMiner(Train)/BigFix} & \multicolumn{3}{c}{BigFix(Train)/CPatMiner} \\
			\cline{2-7}
				& Top-1  & Top-3 & Top-5 & Top-1 & Top-3 & Top-5 \\
			\hline
			Sequencer & 5.4\%&	5.8\%&	6.2\%&	5.3\%	&6.1\%&	7.2\% \\
			CODIT &2.5\%&	4.0\%&	4.4\%&	3.2\%&	5.2\%&	6.4\%\\				
			Tufano19 & 4.5\%&	5.4\%&	5.7\%&	5.9\%&	6.3\%&	7.6\%\\		
			DLFix     & 6.3\%&	6.9\%&	7.3\%&	8.2\%&	8.7\%&	9.2\%\\
			CoCoNuT  & 6.7\% & 7.4\% & 8.1\% & 8.3\% & 9.6\% & 10.7\% \\
                        CURE  & 7.1\% & 7.7\% & 8.2\% & 8.7\% & 9.9\% & 10.9\% \\
			\hline
			{\tool}   & \textbf{7.5\%}&	\textbf{8.1\%}&	\textbf{8.6\%}&	\textbf{9.6\%}&	\textbf{10.2\%}&	\textbf{11.3\%}\\			
			\hline
		\end{tabular}
		\label{RQ2:2}
	\vspace{-2pt}
	\end{center}
\end{table}

Table~\ref{RQ2} shows that {\tool} can fix more bugs than any DL-based
APR baselines on the two large datasets.  Using the top-1 patches,
{\tool} can fix 15.1\% of the total 4,415 bugs in CPatMiner. It fixes
40--322 more bugs than the baselines with top-1 patches.
On BigFix,~it can fix 14.1\% of the total 2,594 bugs
with the top-1 patches. It can fix 31--145 more bugs than those
baselines with the top-1 patches.

Table~\ref{RQ2:2} shows that {\tool} also outperformed the baselines
in the cross-dataset setting in which we trained the models on
CPatMiner and tested them on BigFix and vice versa.



\begin{table}[t]
	\caption{RQ2. Detailed Analysis. Top-1 Result Comparison with DL-based APR Models on CPatMiner Dataset}
	\vspace{-9pt}
	\begin{center}
        \footnotesize
		\renewcommand{\arraystretch}{1}
		\begin{tabular}{p{1.4cm}<{\centering}|p{1.3cm}<{\centering}|p{1.3cm}<{\centering}|p{1.3cm}<{\centering}|p{1.3cm}<{\centering}}
			\hline
		\multirow{2}{*}{\shortstack{\\Types (\#bugs)} }      & CoCoNuT & CURE & DLFix & {\tool} \\
			\cline{2-5}
			
			                    & \#Fixed        & \#Fixed      & \#Fixed & \#Fixed \\\hline
			Type-1 (1,668)      & 28.7\% (479)   &  29.8\%(497)          & 23.7\% (395)    & 29.9\% (499)  \\
			Type-2 (530)        & 1.3\% (7)      & 2.1\% (11)          & 0.6\% (3)    & 4.2\% (22)   \\
			Type-3 (879)        &   12.4\% (109) & 13.2\% (116)  & 11.8\% (104)    & 13.7\% (120)  \\
            Type-4 (1,089)      &  0\% (0)       & 0\% (0)      & 0\% (0)       & 2.0\% (22)   \\
			Type-5 (249)        &  0.4\% (1)     & 0.8\% (2)    & 0.4\% (1)    & 2.0\% (5)  \\
			\hline
			Total (4,415)       &  13.5\% (596)  & 14.2\% (626)  & 11.4\% (503)   & \textbf{15.1\%} \textbf{(667)}  \\
			\hline
		\end{tabular}
		\label{RQ2_3}
	\end{center}
\end{table}

%

Table~\ref{RQ2_3} shows the detailed comparative results on CPatMiner
w.r.t. different bug types.  As seen, {\tool} can {\em auto-fix more
bugs on every type of bug locations on the two large datasets}.
Among 667
fixed bugs, {\tool} has fixed 169 multi-hunk or multi-stmt bugs of
Types 2--5 (i.e., 25.33\% of the total fixed bugs). {\tool} fixes more
bugs (71, 164, and 41 more), and fixes more bugs in each bug
type than the baselines CoCoNuT, DLFix, and CURE, respectively.

{\tool} fixes 52, 61, and 40 more multi-hunk/multi-stmt bugs, and 20,
104, and 2 more one-hunk/one-stmt bugs than CoCoNuT, DLFix, and
CURE. For the multi-statement bugs (Types 2 and 5) that the other
tools fixed, the fixed statements are independent. This result shows that
fixing each individual statement at a time does not work.


\subsection{RQ3. Comparison Results with Pattern-based APR Models}
\label{rq3:sec}

\begin{table}[t]
	\caption{RQ3. Comparison with Pattern-based APR Models}
	\vspace{-10pt}
	\begin{center}
        \small
		\renewcommand{\arraystretch}{1} 
		\begin{tabular}{l|p{0.6cm}<{\centering}p{0.6cm}<{\centering}p{0.6cm}<{\centering}p{0.6cm}<{\centering}p{0.8cm}<{\centering}p{0.6cm}<{\centering}|p{0.5cm}<{\centering}}
			\hline
			Projects  & Chart & Closure & Lang & Math & Mockito & Time & Total\\
			\hline
			ssFix      & 3/7   & 2/11    & 5/12 & 10/26& 0/0     & 0/4  & 20/60\\
			CapGen      & 4/4   & 0/0     & 5/5  & 12/16& 0/0     & 0/0  & 21/25\\
			FixMiner    & 5/8   & 5/5     & 2/3  & 12/14& 0/0     & 1/1  & 25/31\\
			ELIXIR      & 4/7   & 0/0     & 8/12 & 12/19& 0/0     & 2/3  & 26/41\\
			AVATAR      & 5/12  & 8/12    & 5/11 &  6/13& 2/2     & 1/3  & 27/53\\
			SimFix     & 4/8   & 6/8     & 9/13 & 14/26& 0/0     & 1/1  & 34/56\\
			Tbar        & 9/14  & 8/12    & 5/13 & 19/36& 1/2     & 1/3  & 43/81\\
			Hercules    & 8/10  &  8/13     & 10/15 & 20/29 & 0/0     & 3/5  & 49/72\\
			\hline
			
	
			{\bf \tool}   	    & 8/16  & 7/11    & 8/15 & 20/41& 1/2     & 3/6  & \textbf{47}/{\bf 91}\\

			\hline
		\end{tabular}
		
		{\footnotesize X/Y: are the numbers of correct and plausible patches; Dataset: Defects4J}
		\label{RQ3}
	\end{center}
\vspace{-5pt}
\end{table}



As seen in Table~\ref{RQ3}, {\tool} performs at the same level in
terms of the number of bugs as the top pattern-based tools Hercules
and Tbar.

Table~\ref{RQ3_1} displays the details of the comparison
w.r.t. different bug types. As seen, {\tool} fixes 7
Multi/Mix-Statement bugs~(Types 2, 4--5) that Hercules
missed. Investigating further, we found that Hercules is designed to
fix {\em replicated bugs}, i.e., the hunks must have similar
statements.  Those 7 bugs are non-replicated, i.e., the buggy hunks
have different buggy statements or a buggy hunk has multiple
non-similar buggy statements.  For Types 1 and 3, {\tool} fixes 9 less
one-statement bugs~than Hercules due to its incorrect fixes. In total,
{\tool} fixes {\em 12} bugs that Hercules misses: {\em
Chart-7,16,20,24; Time-7; Closure-6,10,40; Lang-10; Math-41,50,91}.



%
Compared to Tbar, {\tool} fixes 15 more multi-hunk/multi-stmt
bugs. Tbar is not designed to fix multi-statements at once as
{\tool}. Instead, it fixes one statement at a time, thus, does not
work well when those 15 bugs require dependent fixes to multiple
statements. The 3 bugs of Type 2 that Tbar can fix are the ones that
the fixes to individual statements are independent. The same reason is
applied to SimFix. Tbar fixes 11 more correct one-hunk/one-statement
bugs.

In brief, we raise {\tool}, a DL-based model, to the~{\em comparable
and complementary} level with those pattern-based APR models.

\begin{table}[t]
	\caption{RQ3. Detailed Comparison with Pattern-based APRs}
	\vspace{-8pt}
	\begin{center}
        \tabcolsep 1pt
		\small
		\renewcommand{\arraystretch}{1} 
		\begin{tabular}{p{4.3cm}<{\centering}|p{1cm}<{\centering}|p{0.6cm}<{\centering}|p{1.1cm}<{\centering}|p{1cm}<{\centering}}
			\hline
			Bug Types & SimFix & Tbar & Hercules & {\tool}\\\hline
			
			Type 1. One-Hunk One-Stmt & 30 & 40 & 34 & 29\\
			Type 2. One-Hunk Multi-Stmts & 1 & 3 & 0 & 4\\ 
			Type 3. Multi-Hunks One-Stmt & 3 & 0 & 15 & 11\\
			Type 4. Multi-Hunks Multi-Stmts & 0 & 0 & 0 & 1\\
			Type 5. Multi-Hunks Mix-Stmts & 0 & 0 & 0 & 2\\\hline
			Total & 34 & 43 & 49 & 47 \\
			\hline
		\end{tabular}
		\label{RQ3_1}
	\end{center}
\vspace{-5pt}
\end{table}


\subsection{RQ4. Sensitivity Analysis}
\label{rq4:sec}


\begin{table}[t]
	\caption{RQ4. Sensitivity Analysis on Defects4J}
	\vspace{-8pt}
	\tabcolsep 3pt
	\small
	\begin{center}
		\renewcommand{\arraystretch}{1}
		\begin{tabular}{l|p{1.6cm}<{\centering}p{1.4cm}<{\centering}p{2cm}<{\centering}|p{1.1cm}<{\centering}}
			\hline
			Variant   & Without Hunk-Det  & Without Expansion & Without Attention-cycle & {\tool}\\
			\hline
			Type-1       & 31      & 30     & 26     & 29  \\
			Type-2       & 4       & 0      & 2      & 4    \\
			Type-3       & 0       & 13     & 9      & 11   \\
			Type-4       & 0       & 0      & 1      & 1   \\		
			Type-5	     & 0       & 0      & 2      & 2    \\
			\hline
			Total    	 & 35      & 43     & 40     & 47   \\
			\hline
		\end{tabular}
		\label{RQ4}
	\end{center}
\vspace{-3pt}
\end{table}



\subsubsection{Impact of Fixing-together Hunk Detection}

As seen in Table~\ref{RQ4}, without hunk detection, \tool can auto-fix
35~bugs. With hunk detection, \tool can fix 14 more multi-hunk bugs
(Types 3-5). It fixes two less Type-1 bugs due to the incorrect hunk
detection. In brief, hunk detection is useful since the
multi-hunk/multi-statement bugs require dependent fixes to multiple hunks
at once.

\subsubsection{Impact of Multi-Statement Expansion}


As seen in Table~\ref{RQ4}, without expansion, {\tool} fixes {\em 43}
bugs in Defects4J. With expansion, it fixes {\em 7} more multi-stmt
bugs in Types 2,4,5 while it fixes two less Type-3 bugs and one less
Type-1 bug. The reason of fixing less bugs in these two types is that
the multi-statement expansion may expand the buggy hunk incorrectly by
regarding a single-statement bug as a multi-statement bug. Even so,
{\tool} still can fix more bugs, showing the usefulness of the
multi-statement expansion.

To compare the impact of Hunk Detection and Multi-Statement Expansion,
 let us note that the variant of {\tool} without Hunk-Detection missed
 all 14 multi-hunk bugs (Types 3,4,5). The variant without Expansion
 missed all 7 multi-statement bugs (Types 2,4,5). However, let us
 consider how challenging it is to fix them. Among 14 multi-hunk bugs
 fixed with Hunk-Detection, 11 bugs are of Type-3
 (multi-hunk/one-statement), in which some approaches (e.g., Hercules)
 can handle by fixing one statement at a time. Only 3 bugs are of Types 4--5. In
 contrast, all 7 bugs fixed with Expansion are multi-statement bugs
 (Types 2,4,5), which cannot be fixed by existing DL-based APR
 approaches. Thus, Expansion contributes to handling more challenging
 bugs than Hunk-Detection.


\subsubsection{Impact of Tree-based LSTM model with Attention and Cycle Training}

(Attention-cycle) To measure the impact of Attention and Cycle
Training, we removed those two mechanisms from {\tool} to produce a
baseline. Our results show that in Defects4J, {\tool} fixes 7 more
bugs on all bug types than the baseline (17.5\% increase). This result
indicates the usefulness of the two mechanisms.



\subsubsection{Impact of Training Data's Size}

Table~\ref{splitting} shows that the size of training data
has impact on {\tool}'s performance.
As seen in Table~\ref{splitting}, the more training data, the higher
the {\tool}'s accuracy. This is expected as {\tool} is a data-driven
approach. But even with less training data (70\%/30\%), {\tool}
achieves 11.7\% for top-1 result, which is still higher than DLFix
(11.4\% in top-1) and Sequencer (7.7\% in top-1); both are with more
training data (90\%/10\% splitting).


\begin{table}[t]
\caption{Impact of the Size of Training Data}
	\vspace{-7pt}
	\tabcolsep 2pt
	\small
	\begin{center}
\begin{tabular}{|c|l|l|l|}
  \hline
  Splitting Scheme on CPatMiner dataset & 90\%/10\% & 80\%/20\% & 70\%/30\% \\
  \hline
  \% Total Bugs at Top-1 & 15.1\% & 13.8\% & 11.7\% \\
  \hline
\end{tabular}
\label{splitting}
	\end{center}
\vspace{-3pt}
\end{table}

\subsection{RQ5. Time Complexity and Parameters}
\label{rq5:sec}

Training time of {\tool} on CPatMiner was +22 hours and predicting on
CPatMiner took 2.4-3.1 seconds for each candidate patch.  Training of
{\tool} on BigFix took 18-19 hours and predicting on BigFix took 3.6-4.2
seconds for each candidate. Predicting on Defects4J took only 2.1
seconds for a candidate due to a much smaller dataset. Test
execution time was +1 second per test case. Test validation took 2--20
minutes for all the test cases for a bug fix.





The best baseline, CURE~\cite{cure-icse21}, fixes fewer bugs than
{\tool} (RQ1 and RQ2), and requires {\em 7 and 7.3 times more}
training parameters than {\tool} on CPatMiner and BigFix,
respectively. Specifically, {\tool} and CURE require 0.39M and 3.1M
training parameters on CPatMiner, and 0.42M and 3.5M parameters on
BigFix. Thus, {\tool} is less complex than CURE, while achieving
better results.


\subsubsection*{Threats to Validity}

We tested on Java code. The key modules in {\tool} are
language-independent, except for the third-party FL and post-processing
with program analysis.
%
Pattern-based APR tools require a dataset with test cases, thus, we
compared them on Defects4J only. We tried our best to re-implement the
pattern-based APR baselines and CURE for
a fair comparison.

\subsubsection*{Illustrative Example}

\begin{figure}[t]
	\centering
	\lstset{
		numbers=left,
		numberstyle= \tiny,
		keywordstyle= \color{blue!70},
		commentstyle= \color{red!50!green!50!blue!50},
		frame=shadowbox,
		rulesepcolor= \color{red!20!green!20!blue!20} ,
                xleftmargin=1.5em,xrightmargin=0em, aboveskip=1em,
		framexleftmargin=1.5em,
                numbersep= 5pt,
		language=Java,
		language=Java,
		basicstyle=\footnotesize,
                moredelim=**[is][\color{red}]{@}{@},
		escapeinside= {(*@}{@*)}
	}
	\begin{lstlisting}
public void excludeRoot(String path) {
@-   String url = toUrl(path);@
@-   findOrCreateContentRoot(url).addExcludeFolder(url);@
(*@{\color{cyan}{+\quad Url url = toUrl(path);}@*)
(*@{\color{cyan}{+\quad findOrCreateContentRoot(url).addExcludeFolder(url.getUrl());}@*)
}
public void useModuleOutput(String production, String test) {
    modifiableRootModel.inheritCompilerOutputPath(false);
@-   modifiableRootModel.setCompilerOutputPath(toUrl(production));@
@-   modifiableRootModel.setCompilerOutputPathForTests(toUrl(test));@
(*@{\color{cyan}{+\quad modifiableRootModel.setCompilerOutputPath(toUrl(production).getUrl());}@*)
(*@{\color{cyan}{+\quad modifiableRootModel.setCompilerOutputPathForTests(toUrl(test).getUrl());}@*)
}
	\end{lstlisting}
	\vspace{-16pt}
	\caption{A Multi-hunk/Multi-statement Fix in CPatMiner}
	\label{case1}
	\vspace{-7pt}
\end{figure}


Figure~\ref{case1} shows a correct fix from {\tool}. It correctly
detects two buggy hunks; each with multiple statements.
%
%
{\tool} leverages the variable names existing in the same method
(\code{modifiableRootModel} at line 8) in composing the fixed code at
lines~11--12.
The DL-based baselines, Sequencer~\cite{chen2018sequencer} and
CoCoNuT~\cite{lutellier2020coconut},~treat code as sequences, and does
not derive well the structural changes for this fix. DLFix fixes one
statement at a time, thus, does not~work (the fixes at line 2 and line 3
depend on each other). For pattern-based
APRs~\cite{hercules-icse19,Simfix}, there is no fixing template for
this bug.

\subsubsection*{Limitations}

{\tool} has the following limitations. First, as with ML approaches,
fixes with rare or out-of-vocabulary names are challenging. With more
training data, {\tool} has higher chance to encounter the ingredients
to generate a new name. Second, we focus only on the bugs that cause
failing tests. Security, vulnerabilities, and non-failing-test bugs
are still its limitations. Third, we cannot generate fixes with several
new statements added or arbitrarily large sizes of dependent fixed
statements.
Fourth, the expansion algorithm produces incorrect hunks to be
fixed, leading to fixing incorrect statements.
Finally, we currently focus on Java, however, the basic
representations used in {\tool}, e.g., token, AST, dependency, are
universal to any program language. Only third-party FL and
post-processing with semantic checkers are language-dependent.


\section{Related Work}
\label{related:sec}

\vspace{2pt}
\noindent{\bf Deep Learning-based APR approaches}. 
DeepRepair~\cite{white2016deep} learns code similarities to select the
repair ingredients from code fragments similar to the buggy code.
DeepFix~\cite{gupta2017deepfix} learns the syntax rules to fix syntax
errors.
Ratchet~\cite{hata2018learning}, Tufano {\em et
al.}~\cite{tufano2018empirical}, and
SequenceR~\cite{chen2018sequencer} mainly use neural network machine
translation (NMT) with attention-based encoder-decoder and code
abstractions to generate patches.
CODIT~\cite{chakrabortycodit} encodes code structures, learns code
edits, and adopt an NMT model to suggest fixes.
Tufano {\em et al.}~\cite{tufano2019learning} learn code changes
using a sequence-to-sequence NMT with code abstractions and keyword
replacing. DLFix~\cite{icse20} has a tree-based translation model to
learn the fixes.  CoCoNuT~\cite{lutellier2020coconut} develops a
context-aware NMT model. CURE~\cite{cure-icse21} proposes a
code-aware NMT using GPT model~\cite{radford2018improving}.
%
The existing DL-based APR models fix individual statements at a
time and are ineffective for multi-hunk/multi-stmt bugs.

\vspace{3pt}
\noindent{\bf Pattern-based APR approaches.}
Those approaches have {\em mined and learned fix patterns}
from prior fixes
\cite{nguyen2013semfix,le2016history,liu2019avatar,kim2013automatic},
either automatically or
semi-automatically~\cite{le2016history,nguyen2013semfix,liu2019avatar,tbar-issta19}.
Prophet~\cite{long2016automatic} learns code correctness models from a
set of successful human patches.  Droix~\cite{tan2018repairing} learns
common root-causes for crashes using a search-based repair.
Genesis~\cite{long2017automatic} automatically infers patch generation
from users' submitted patches. HDRepair~\cite{le2016history} mines fix
patterns with graphs. ELIXIR~\cite{saha2017elixir} uses templates from
PAR with local variables, fields, or constants, to build
fixed expressions. CapGen~\cite{wen2018context},
SimFix~\cite{Simfix}, FixMiner~\cite{koyuncu2018fixminer} rely on
frequent code changes extracted from existing
patches. Avatar~\cite{liu2019avatar} exploits fix patterns of static
analysis violations.
Tbar~\cite{tbar-issta19} is a template-based APR tool with the
collected fix patterns.
Angelix~\cite{mechtaev2016angelix} catches program semantics to
fix methods.
ARJA~\cite{yuan2018arja} generates lower-granularity patch
representation enabling efficient searching. We did not
compare with Angelix since we compared with CapGen that outperforms
Angelix. We could not reproduce ARJA, however, ARJA fixes
only 18 bugs, while {\tool} fixes 42 bugs on the same four projects
in~Defects4J.

\section{Conclusion}
\label{conclusion:sec}


In this work, we make three key contributions: 1) a novel FL technique
for multi-hunk, multi-statement fixes combining traditional SBFL
with deep learning and data-flow analysis; 2) a compositional approach to
generate multi-hunk, multi-statement fixes with
divide-and-conquer strategy; and 3) enhancements and
orchestration of a two-layer LSTM model with the attention layer and
cycle training.
%
%
On Defects4J, {\tool} outperforms the DL-based APR baselines from
42\%–683\% in terms of the number of fixed bugs. On BigFix, it fixes
31–145 more bugs with the top-1 patches. On CPatMiner, it fixes
40--52 more multi-hunk/multi-stmt bugs than the baselines.
%





\section*{Acknowledgments}
This work was supported in part by the US National Science Foundation
(NSF) grants CNS-2120386, CCF-1723215, CCF-1723432, TWC-1723198,
CCF-1518897, and CNS-1513263.


\newpage

\balance

\bibliographystyle{ACM-Reference-Format}

\bibliography{sections/Reference}

\end{document}